\documentclass[twocolumn, aps, prb, superscriptaddress]{revtex4-1}

\usepackage{graphicx}
\usepackage{amsmath,amssymb,mathtools,hyperref}
\usepackage{color}

\newcommand{\ket}[1]{\left|#1\right\rangle}
\newcommand{\bra}[1]{\left\langle #1\right|}
\newcommand{\mean}[1]{\left\langle #1\right\rangle}
\newcommand{\opc}[1]{{\hat{c}^{\phantom \dagger}}_{#1}}
\newcommand{\opcdag}[1]{{\hat{c}^{\dagger}}_{#1}}

\begin{document}

\title{Resilience of hidden order to symmetry-preserving disorder} 

\author{Marcello~Calvanese~Strinati}
\affiliation{NEST, Scuola Normale Superiore and Istituto Nanoscienze-CNR, 56126 Pisa, Italy}
\affiliation{D\'epartement de Physique, Ecole Normale Sup\'erieure / PSL Research University,
CNRS, 24 rue Lhomond, 75005 Paris, France}
\affiliation{{Department of Physics, Bar-Ilan University, 52900 Ramat-Gan, Israel}}

\author{Davide~Rossini}
\affiliation{{Dipartimento di Fisica, Universit\`a di Pisa and INFN, Largo Pontecorvo 3, 56127 Pisa, Italy}}

\author{Rosario~Fazio}
\affiliation{ICTP, Strada Costiera 11, 34151 Trieste, Italy}
\affiliation{NEST, Scuola Normale Superiore and Istituto Nanoscienze-CNR, 56126 Pisa, Italy}

\author{Angelo~Russomanno}
\affiliation{NEST, Scuola Normale Superiore and Istituto Nanoscienze-CNR, 56126 Pisa, Italy}
\affiliation{ICTP, Strada Costiera 11, 34151 Trieste, Italy}

\date{\today}

\begin{abstract}
We study the robustness of non-local string order in two paradigmatic disordered spin-chain models, 
a spin-1/2 cluster-Ising and a spin-1 XXZ Heisenberg chain.
In the clean case, they both display a transition from antiferromagnetic to string order.
Applying a disorder which preserves the Hamiltonian symmetries, we find that the transition persists in both models. In the disordered cluster-Ising model we can study the transition analytically -- by applying the strongest coupling renormalization group -- and numerically -- by exploiting integrability to study the antiferromagnetic and string order parameters. We map the model into a quadratic fermion chain, where the transition appears as a change in the number of zero-energy edge modes. {
We also explore its zero-temperature-singularity behavior and find a transition from a non-singular to a singular region, at a point that is different from the one separating non-local and local ordering.} The disordered Heisenberg chain can be treated only numerically: by means of MPS-based simulations, we are able to locate the existence of a transition between antiferromagnetic and string-ordered phase, through the study of order parameters. Finally we discuss possible connections of our findings with many body localization. 
\end{abstract}


\maketitle

\section{Introduction}
\label{sec:Intro} 

Since Landau, we know that 
phase transitions and symmetry breaking are strictly connected (see for instance Ref.~\onlinecite{Goldenfeld:book}). 
Let us focus on zero-temperature quantum phase transitions (QPTs)~\cite{Sachdev:book}. 
In the ordered phase, there is a manifold of degenerate ground states which have high entanglement and obey the same symmetries of the Hamiltonian. Nevertheless, any physical ground state breaks this symmetry: because of decoherence, the system always ends up in a superposition of these symmetry-preserving states which has small entanglement and breaks the symmetry. This mechanism manifests in the expectation value $\Phi$ of some local operator $\hat{\Phi}({\bf x})$ (the order parameter) being different from zero. Symmetry breaking always comes together with long-range order: in the thermodynamic limit, the order parameter in a ground state is infinite-range correlated, 
$\lim_{|{\bf x}-{\bf y}|\to\infty} \big\langle \hat{\Phi}^\dagger({\bf x})\hat{\Phi}({\bf y}) \big\rangle_{\rm GS}=|\Phi|^2$. 
Although the value of $\Phi$ depends on the specific choice of the symmetry-breaking ground state, 
its modulus $|\Phi|^2$ is independent of it. Moreover, this correlator does not depend on the choice of the ground state, even if we consider a non-physical symmetry-preserving ground state. From a physical point of view, the order parameter can be, for instance, the magnetization in the ferromagnetic transition (breaking of rotation symmetry) or the superconducting wave-function in superconductivity (breaking of gauge symmetry). 

This paradigm has been challenged in the last decades by the discovery of topological phase transitions which
are characterized by no local order parameter but by a global rearrangement of the system structure~\cite{Bernevig:book, Simons:book}.
The landscape of topological phase transitions is extremely rich~\cite{Nobel_lecture}; here we specifically focus on a particular class of topological 
phases occurring in one-dimensional systems,
for which the concept of hidden order has been put forward. In this case, the order parameter 
still exists, but it is a non-local one: it is a string operator involving the system in its globality. 
The most famous example of string order (SO) is in the Haldane phase of a spin-1 isotropic Heisenberg chain~\cite{Haldane_1983}. 
Let us focus on one-dimensional systems and call $\hat{\mathcal{O}}_{j,k}$ the SO parameter 
between two sites $j$ and $k$: long-range order in the thermodynamic limit is given by 
$\lim_{|j-k|\to\infty} \big\langle \hat{\mathcal{O}}_{j,k} \big\rangle_{\rm GS}\neq0$,
where $\langle \, \cdot \, \rangle_{\rm GS}$ denotes the ground-state (GS) average. 
In this case, an infinite-range {\em non-local} correlator is different from zero and does not correspond 
to any local non-vanishing object. 
Examples of transitions to hidden SO phases are the one between a ferromagnet 
and the Haldane phase in spin-1 chains~\cite{Sanctuary_2003, DegliEsposti_2004, Ueda_2008}, 
and the one occurring in the extended Bose Hubbard model~\cite{DallaTorre_2006, Berg_2008, Rossini_NJ2012}. 
In some cases of hidden order the non-local parameter has not yet been recognized or does not exist~\cite{Erez_PRB10,Emanuele_arXiv17}, 
and recently SO has been discovered in a periodically driven spin-1/2 chain~\cite{Emanuele_arXiv16}.

Properties of systems undergoing QPTs are markedly altered by the presence of disorder. 
Disorder shifts the phase transition point and can also create new phases, 
like the Griffiths phase~\cite{Griffiths, Mohit, Bray, Boechat} in disordered ferromagnets, 
the insulating phase in disordered superconductors~\cite{Yona}, the Bose-Glass phase 
in a disordered Bose-Hubbard model~\cite{MFisher_PRB} and the many body localized phase in short-range
interacting quantum systems~\cite{Basko_Ann06,Oganesyan_PRA75}. 
Very remarkable are the works by Dasgupta and Ma~\cite{Dasgupta} and by Fisher~\cite{Fisher_1995}, 
who are able to construct a renormalization group (RG) flow -- the ``strongest coupling renormalization group'' --
to understand phase transitions in such models. {This RG method has been used afterwards for a large variety of random quantum and classical systems (see Ref.~\onlinecite{Igloi_review} for a review)}. Many works focused on disorder and phases 
with local order. The interplay of disorder and non-local SO has been considered in a comparatively small amount of papers. {Ref.~\onlinecite{Igloi} focused on disordered spin-1/2 ladders, that may describe properties of spin-1 chains with SO~\cite{Scalapino}. 
The interest on the ladders was due to the fact that the application of strongest coupling RG to spin-1 chains is difficult due to the proliferation of large local spin terms. This difficulty has been overcome in Refs.~\onlinecite{Boechat1,Boechat,Lajko_PRB}; in particular, 
Ref.~\onlinecite{Boechat} found a transition from a Griffiths to a disordered Haldane phase in an $S=1$ disordered
  antiferromagnetic Heisenberg chain. The numerical demonstration of the persistence of the string order in this model (with a different disorder) and a detailed analysis of the zero-temperature-singularity behaviour has been discussed in Ref.~\onlinecite{Lajko_PRB}. In these works, the form of the disorder has been chosen so to preserve the symmetries of the Hamiltonian. 
 While for small disorder amplitudes the SO is preserved thanks to the topological protection,
  over a given threshold the non-local ordering is broken.
  The topological protection for small disorder is expected from general arguments, however
  the breaking of SO for strong disorder and the properties of this transition are not trivial at all~\cite{Lajko_PRB}.}

{As far as we know, there are no studies on the effect of a strong disorder on the properties of a transition from a string non-local order
to a local order (like ferro- or antiferro- magnetic).}
Here, we make a progress in this direction considering disordered spin-1 and spin-1/2 chains 
whose clean counterparts (the XXZ Heisenberg chain and the cluster-Ising model) are well known to display 
a transition from an antiferromagnetic (AF) to a string-ordered phase (the isotropic Heisenberg chain 
is the Haldane model). 
In the first case we resort to numerics: using a variational algorithm based on the formalism
of matrix-product states (MPS)~\cite{Schollwock2011} we find that a phase with SO actually does exist, 
even in the disordered model. 
We locate the phase transition point and observe that it is shifted by the disorder, with respect to the clean case. 
In the case of the disordered cluster-Ising model we can go further: by applying 
a strongest coupling RG transformation, we show the existence of an AF and a string-ordered phase, 
and analytically find the phase transition point. 
In this case, the model is solvable because it can be mapped to a disordered non-interacting fermion chain. 
In the fermionic representation, the transition is of topological nature and is 
characterized by the appearance of zero-energy edge modes.
{Finally, we 
  study the zero-temperature-singularity behavior of the model, by numerically analyzing the properties
  of the logarithmic gap distribution and of the inverse average of the gap~\cite{Boechat,Lajko_PRB}.
  We find a transition between a non-singular and a singular behavior when the disorder strength is increased.
  This transition point is different from the one between SO and AF: for our specific form of disorder, we see the existence
  of a non-singular and a singular SO phase, similarly to the findings of Ref.~\onlinecite{Lajko_PRB}.
  Moreover we observe that the AF phase is singular.}

The paper is organized as follows. In Sec.~\ref{sec:Models}, we introduce the two models we are considering. 
In Sec.~\ref{CIM:sec}, we study the AF/SO transition in the cluster-Ising chain. 
We do this both analytically (through the strongest coupling RG) and numerically (evaluating the appropriate 
correlators thanks to the free-fermion mapping); we also study the properties of edge modes 
in the fermionic representation. 
In Sec.~\ref{XXZ:sec}, we study the AF/SO transition in the Heisenberg XXZ chain, 
by resorting to MPS numerical simulations. 
Finally, in Sec.~\ref{conco:sec}, we draw our conclusions. 
In Appendix~\ref{app:RG}, we show details on the strongest coupling RG transformation applied 
to the cluster-Ising chain, and in Appendix~\ref{sec:appendixdetailsonthekinks} we discuss the appearance 
of kinks in the MPS approximation of the GS, which are due to a numerical artifact.

\section{The models}
\label{sec:Models}

In the following, we will consider two spin chains that exhibit a zero-temperature QPT
between a N\'eel-like AF phase and a phase that is characterized by non-local SO. They are the spin-1/2 cluster-Ising model
and the spin-1 XXZ Heisenberg chain. The two models enjoy a local $\mathbb{Z}_2$ and $\mathbb{D}_2$ symmetry, respectively. We thus expect the SO phase to exist also in the presence of disorder, as long as the local symmetry is not broken~\cite{Pollmann_2010,PerezGarcia_2008}. In this section, we introduce the two models by presenting their Hamiltonian together with some of their properties and the relevant antiferromagnetic and string order parameters.

\subsection{Spin-1/2 cluster-Ising model} \label{CIM_mod:sec}

The simplest and exactly solvable model in this context is the so called cluster-Ising model (CIM)~\cite{Smacchia_2011, Son_2011},
which is described by the Hamiltonian
\begin{equation}
  \hat{H}_{\rm CIM} = \sum_j \big[ - J_j \hat{\sigma}_{j-1}^x\hat{\sigma}_j^z\hat{\sigma}_{j+1}^x
  + \lambda_j \hat{\sigma}_j^y\hat{\sigma}_{j+1}^y \big] \,.
  \label{eq:CIM}
\end{equation}
Here, $\hat \sigma^\alpha_j$ (with $\alpha = x,y,z$) denote the spin-1/2 Pauli matrices on the $j$-th site of the chain
($j=1, \ldots, L$, where $L$ is the chain length),
while $J_j$ and $\lambda_j$ denote the (possibly site-dependent) three-spin and two-spin coupling terms, respectively.
In the following, we will always adopt open boundary conditions (OBC).
After a standard Jordan-Wigner transformation~\cite{Lieb_AP61} of the form
$\hat c_j = \big( \prod_{m=1}^{j-1} \hat \sigma^z_m \big) \hat \sigma^-_j/2$,
the Hamiltonian in Eq.~\eqref{eq:CIM} can be mapped onto a free-fermion model with nearest-
and next-to-nearest-neighbor hopping terms
\begin{align}
    \hat{H}_{\rm CIM} = \sum_j \big[ & - J_j (\opcdag{j-1}-\opc{j-1})(\opcdag{j+1}+\opc{j+1})\nonumber\\
  &+ \lambda_j (\opcdag{j}+\opc{j})(\opcdag{j+1}-\opc{j+1}) \big] \,,
  \label{eq:CIM_fermion}
\end{align}
in a way similar to the one-dimensional quantum Ising model in transverse field~\cite{Pfeuty_AP70}. 
This Hamiltonian is integrable, and its dynamics can be reduced to that of non-interacting 
fermionic quasiparticles: the GS is the BCS state without quasiparticles.

It has been proven~\cite{Smacchia_2011} that, in the homogeneous case (i.e. $J_j=J$ and $\lambda_j = \lambda$ for all $j$),
the system features a QPT between a conventional AF phase along the $y$ direction and a phase with 
non-local SO (the so called ``cluster phase''), when decreasing $\lambda/J$ and crossing the critical value of $1$.

The $y$-antiferromagnet is detected by a local order parameter. In agreement with the discussion 
in the Introduction, we can express it as an infinite-range correlator:
\begin{equation}
  \big( \mathcal{S}_{[1/2]}^{y} \big)^2 = \frac{1}{4}\,{\lim_{l\to\infty} (-1)^l \, 
    \big\langle \hat{\sigma}_k^y \hat{\sigma}_{k+l}^y \big\rangle} \,.
  \label{eq:neelCIM}
\end{equation}
To be precise, here we are considering the square modulus of the order parameter, 
which is the same for all the degenerate symmetry-breaking ground states. 
Since this model breaks the $\mathbb{Z}_2$ symmetry (see below for more details), there are exactly 
two symmetry-breaking ground states which differ for the sign of the order parameter.
Moreover, because we are evaluating the modulus of the order parameter as the limit of a correlator, 
it is not important to select the symmetry breaking ground states: the correlators are the same 
on all the states of the GS manifold. From a technical point of view, the mapping 
to the fermionic model is very important in order to evaluate this correlator. 
Thanks to Wick's theorem, it can be expressed as a Toeplitz determinant~\cite{Barouch_PRA71} 
of specific single-particle fermionic correlators 
of the Hamiltonian~\eqref{eq:CIM_fermion} -- see Subsec.~\ref{CIM_num:sec} for more details. 

On the opposite, the cluster phase is characterized by a non-vanishing value of the non-local SO parameter, which is expressed as the infinite-range limit of a non-local correlator:
\begin{equation} 
  \mathcal{O}_{[1/2]}^z = \! \lim_{l \to \infty} (-1)^l \, \big\langle \hat{\sigma}_k^x\hat{\sigma}_{k+1}^y 
  \bigg[ \prod_{n=k+2}^{k+l-2} \!\! \hat{\sigma}_n^z \bigg] \hat{\sigma}_{k+l-1}^y \hat{\sigma}_{k+l}^x \big\rangle \,.
  \label{eq:stringCIM}
\end{equation}
From a technical point of view, this non-local correlator is evaluated by applying to the Hamiltonian 
a duality transformation (see Ref.~\onlinecite{Smacchia_2011} and Appendix~\ref{app:RG} for details). 
This non-local unitary transformation maps the Hamiltonian onto itself, with $\lambda_j$ exchanged with $J_j$ 
[see Eq.~\eqref{Ham:duale:eqn}]. Moreover, the correlator of Eq.~\eqref{eq:stringCIM} 
is mapped on the AF correlator of Eq.~\eqref{eq:neelCIM}, which can be evaluated as explained above.

Unless specified, here and in the following equations, we will omit the subscript $\langle \, \ldots \, \rangle_{\rm GS}$ 
and consider all the expectation values on one of the degenerate ground states of the system. 
As remarked before, it is not important which GS in particular, 
being the correlators independent of the specific choice of the state in the GS manifold. 
Moreover, in the presence of disorder, we will average over many realizations: 
this operation will be denoted by an overline $\overline{\mean{\ldots}}$. 
We notice that the disorder is not translationally invariant: 
translation invariance is restored after averaging over the disorder realizations.
From a numerical point of view, we can only evaluate finite range correlators: 
${\mathcal{O}}_{[1/2],l}^\alpha$ (with $\alpha=x,y,z$) denotes the string correlator over a range $l$,
such that $\mathcal{O}_{[1/2]}^\alpha = \lim_{l \to \infty} {\mathcal{O}}_{[1/2],l}^\alpha$.
In a similar way, we define the finite-range AF correlator as
$\big( \mathcal{S}_{[1/2],l}^\alpha \big)^2$ such that
$\big( \mathcal{S}_{[1/2]}^\alpha \big)^2 = \lim_{l \to \infty} \big( \mathcal{S}_{[1/2],l}^\alpha \big)^2$. 

Let us now focus on the version of the Hamiltonian with disorder, which we are studying in the rest of the work.
A necessary condition to preserve the string-ordered phase is the choice of a disorder that preserves 
the symmetry of the model: this is the symmetry broken by the GS 
in the phase transition. As mentioned above, the Hamiltonian in Eq.~\eqref{eq:CIM} enjoys a $\mathbb{Z}_2$ symmetry, 
since it is invariant under a rotation of angle $\pi$ around the $z$ axis:
$\hat{V}^\dagger\hat{H}_{\rm CIM}\hat{V} = \hat{H}_{\rm CIM}$ with 
$\hat{V} = \exp \big(-i\frac{\pi}{2}\sum_j\hat{\sigma}_j^z\big)$.
Being the symmetry group generated by only two operators ($\hat{V}$ and $\hat{\mathbb{I}}$), the GS manifold
can have at most dimension two, so there can be at most two symmetry-breaking ground states.
The invariance under the operator $\hat{V}$ implies:
\begin{equation}
  \Big[ \prod_j\hat{\sigma}_j^z,\hat{H}_{\rm CIM} \Big]=0\,.
\end{equation}
A possibility to satisfy this condition in a disordered situation is to assume that both the site-dependent two-spin and
the three-spin couplings are taken randomly from some probability distribution. In the following, we will specifically address the situation in which
both the $\lambda_j$ and the $J_j$ are uniformly distributed over some interval
$\lambda_j \in [0,\lambda_{\rm max}]$ and $J_j \in [0, J_{\rm max}]$, for all $j=1, \ldots, L$.

\subsection{Spin-1 XXZ Heisenberg chain} \label{Heis:XXZ:subsec}

The other model that we are going to focus on is a spin-1 XXZ Heisenberg chain, given by the Hamiltonian
\begin{equation}
  \hat H_{\rm XXZ} = J \sum_j \left[ \hat S^x_j\hat S^x_{j+1}+\hat S^y_j\hat S^y_{j+1}+{\Lambda}_j\hat S^z_j\hat S^z_{j+1} \right] \, ,
  \label{eq:XXZ}
\end{equation}
where $\hat S_j^\alpha$ (with $\alpha = x,y,z$) now denotes the spin-1 operators on the $j$-th site, $J$ is the energy scale
of the nearest-neighbor spin coupling, and ${\Lambda}_j$ is the anisotropy factor along the $z$ axis, at site $j$.
Contrary to the CIM, the spin-1 Heisenberg chain is a non-integrable model and cannot be easily diagonalized.
For this reason, numerical approaches based on exact diagonalization techniques or on MPS
are usually employed in order to capture the GS physics.
In the clean case, that is for ${\Lambda}_j = {\Lambda}, \; \forall j$, model~\eqref{eq:XXZ} is known to display 
a phase transition between a topological phase, usually referred to as the Haldane phase~\cite{Haldane_1983}, 
and a non-topological N\'eel AF phase.
Such transition has been studied in some details in the literature, and is expected to occur
for ${\Lambda} \approx 1.186\ldots$~\cite{DegliEsposti_2004, Ueda_2008}.

In an open chain, the GS in the Haldane phase is identified by the presence of gapless spin-1/2 modes on top of a gapped bulk, which make the GS
four-fold degenerate in the thermodynamic limit. This phenomenology is related to a hidden $\mathbb{D}_2$ symmetry breaking, described by the dihedral group of rotations~\cite{Pollmann_2010, Pollmann_2012}
\begin{equation}
  \mathcal{G}_{\mathbb{D}_2} = \big\{\mathbb{\hat I},e^{i\pi\sum_n\hat S^x_n},e^{i\pi\sum_n\hat S^y_n},e^{i\pi\sum_n\hat S^z_n} \big\} \, .
\end{equation}
Because of the presence of such edge modes, a state in the Haldane phase is characterized by a hidden long-range order~\cite{DenNijs_1989},
which cannot be revealed by any expectation value of simple two-point correlators
$\langle \hat S^\alpha_k\hat S^\alpha_{k+l} \rangle$.
Indeed, their GS expectation values vanish in the limit $l \to \infty$.
This hidden order can be seen by defining a non-local SO parameter for a spin-1 chain 
in a way similar to what has been done for the CIM~\eqref{eq:stringCIM}:
\begin{equation}
  \mathcal{O}^\alpha_{[1]} = \lim_{l \to \infty}{\mathcal{O}}^\alpha_{[1],l} =  \lim_{l \to \infty} \big\langle \hat S^\alpha_k \bigg[\,\prod_{n=k+1}^{k+l-1}e^{i\pi\hat S^\alpha_n}\,\bigg] \hat S^\alpha_{k+l} \big\rangle \,,
\label{eq:SO-1}
\end{equation}
such that the system is said to posses SO if the above limit is finite and nonvanishing.
The presence of hidden order can be understood by remapping the model onto a ferromagnetic chain 
with four symmetry-broken states,
by means of the Kennedy-Tasaki transformation~\cite{Kennedy_1992}: string correlators are mapped
onto two-point correlators, which indeed reveal the presence of ferromagnetic order. 
This is similar to what happens in the CIM, where a duality transformation maps the SO correlator 
of Eq.~\eqref{eq:stringCIM} onto the AF correlator of Eq.~\eqref{eq:neelCIM}. 

The physical meaning of the SO parameter~\eqref{eq:SO-1} is very clearly expressed in Ref.~\onlinecite{Scalapino}; 
we briefly review it here for reader's convenience: if we measure the value of the spin projections along $\alpha$, 
we can find 1,0 or $-1$, being this a spin-1 chain. Since the GS is a superposition of spin eigenstates, 
each time we measure we find a different sequence of $+1$,0 or $-1$, with some probability. 
The fact that the operator~\eqref{eq:SO-1} has a non-vanishing expectation value means that, if we withdraw the zeros 
from any of these sequences, we find alternatively $+1$ and $-1$: without the zeros, the system behaves 
antiferromagnetically. This property cannot be witnessed by any local operator: only the non-local string operator can do.

On the other hand, the presence of long-range AF order along the $z$ axis in the N\'eel phase is witnessed 
by a nonzero value of the staggered two-point correlator 
\begin{equation}
  \big(\mathcal{S}_{[1]}^\alpha\big)^2 = \lim_{l \to \infty} \big( \mathcal{S}_{[1],l}^\alpha \big)^2 
  = \lim_{l \to \infty} (-1)^l \, \big\langle \hat S^\alpha_k \hat S^\alpha_{k+l} \big\rangle \,,
\label{eq:AF-1}
\end{equation}
with $\alpha = z$ (this is the square value of the AF order parameter along $\alpha$).
Thus, we can identify the Haldane phase of the XXZ chain by a nonzero SO [Eq.~\eqref{eq:SO-1}, for all $\alpha$]
and a vanishing expectation value of the staggered correlator in Eq.~\eqref{eq:AF-1}, for any $\alpha$.
Conversely, the N\'eel phase is identified by a vanishing SO $\mathcal{O}^\alpha_{[1]}$ for $\alpha=x,y$,
and by a nonzero value of the staggered AF order parameter ${\mathcal{S}^z_{[1]}}$.
We point out that we cannot use the SO along $z$ as an order parameter of the Haldane phase,
because the observable $\mathcal{O}^z_{[1]}$ is nonzero both in the Haldane and in the N\'eel phase~\cite{Su_2012}.
This can be simply understood in the large-${\Lambda}$ limit, where the GS is given by a product
of consecutive states with opposite spin projection. 
From Eq.~\eqref{eq:SO-1}, it is clear that $\hat{\mathcal{O}}^z_{[1],l}$ evaluated on such GS is exactly $-1$
for all $l$, and in particular for $l \to \infty$.

We will model the disorder by taking ${\Lambda}_j$ as a random variable,
which is uniformly distributed between ${\Lambda}_{\rm min}$ and ${\Lambda}_{\rm max}$.
We stress that a necessary condition for the GS to possess SO is to enjoy a unitary and local symmetry~\cite{PerezGarcia_2008}. 
As stated before, since the presence of non-uniform anisotropy ${\Lambda}_j$
does not break the $\mathbb{D}_2$ symmetry, the SO phase is expected to be present also in the disordered
XXZ model, at least for some range of values ${\Lambda}_{\rm min}$ and ${\Lambda}_{\rm max}$ {(see Appendix~\ref{sec:xxzmodelwithmagneticfield} for an example of destruction of string order when a symmetry breaking term is added to the Hamiltonian)}.

\section{Antiferromagnetic -- cluster phase transition in the disordered CIM} \label{CIM:sec}
\subsection{Strongest coupling RG approach} \label{strongest:sec}
To understand whether the CIM can undergo a QPT or not, we perform a strongest coupling RG analysis 
in the thermodynamic limit $L \to \infty$, very similar in spirit to the one used in Ref.~\onlinecite{Fisher_1995} 
for the disordered transverse field Ising model. We consider the largest value of the coupling
\begin{equation}
  \Omega_I=\max\left\{J_j,\,\lambda_j\right\}\,.
\end{equation}
The idea is to diagonalize the part of the Hamiltonian related to this coupling, 
assuming that it is so large that the corresponding term of the Hamiltonian can be considered as non-interacting, in a crudest approximation. The rest of the chain
can be considered as a perturbation which changes the GS energy of this subsystem at second order in the coupling. 
At the end of the renormalization step, we take only the perturbed GS of the renormalized subsystem, 
and discard the rest of its local Hilbert space. In this way, at each renormalization step, we reduce 
the energy scale at which we are looking at the system: the considered site is renormalized away 
and an effective low-energy coupling term is generated. After many applications of the renormalization step, 
we asymptotically reach the GS and we can find its properties.
One can distinguish between two possible cases, depending whether the largest coupling is a given $J_j$, 
or a given $\lambda_j$. We provide the derivation in full detail in Appendix~\ref{app:RG}; 
here, we focus on its main points and their physical meaning.

We start assuming that the largest coupling is $J_j$: the part of the Hamiltonian corresponding to it is given by
\begin{equation}
  \hat{H}_0 = -J_j\hat{\sigma}_{j-1}^x\hat{\sigma}_{j}^z\hat{\sigma}_{j+1}^x \,,
  \label{eq:Hloc_RG}
\end{equation}
while the coupling to the rest of the system can be described by the following operator:
\begin{equation}
  \hat{V}=\lambda_{j-1}\hat{\sigma}_{j-2}^y\hat{\sigma}_{j-1}^y+\lambda_{j+1}\hat{\sigma}_{j+1}^y\hat{\sigma}_{j+2}^y \,.
  \label{eq:V_RG}
\end{equation}
As detailed in Appendix~\ref{app:RG}, it is possible to treat the term $\hat V$ perturbatively, 
applying a first-order perturbation theory to the four-fold degenerate GS of $\hat H_0$.
After diagonalizing the resulting second-order perturbation matrix,
we project over one of the perturbed ground states, ending up into eliminating site $j$ and generating a new coupling
\begin{equation}
  -\widetilde{\lambda}_j\hat{\sigma}_{j-2}^y\hat{\sigma}_{j+2}^y\quad{\rm with}\quad\widetilde{\lambda}_j\simeq \frac{\lambda_{j-1}\lambda_{j+1}}{J_j}\,.
  \label{eq:renormalizedcouplingsrgsigmay}
\end{equation}
The operators $\hat{\sigma}_{j-2/j+2}^y$ in Eq.~\eqref{eq:renormalizedcouplingsrgsigmay} are in principle different from the unrenormalized ones:
they coincide with them up to quartic terms in $\lambda/J$.

In the opposite case, where the largest coupling is $\lambda_j$, one can apply to Eq.~\eqref{eq:CIM} 
a duality transformation
\begin{equation} \label{duality0:eqn}
  \hat{\mu}_j^x = \prod_{k=1}^j\hat{\sigma}_k^z \,, \qquad
  \hat{\mu}_j^z = \hat{\sigma}_j^x\hat{\sigma}_{j+1}^x\,,
\end{equation}
which maps the Pauli operators $\hat{\sigma}_j^\alpha$ onto different Pauli operators $\hat{\mu}_j^\alpha$. 
After the application of this transformation, the CIM Hamiltonian in terms of $\hat{\sigma}_j^\alpha$ [Eq.~\eqref{eq:CIM}] is re-expressed in terms of $\hat{\mu}_j^\alpha$ [see Eq.~\eqref{Ham:duale:eqn} in Appendix~\ref{app:RG}]: the transformed Hamiltonian has the same form of Eq.~\eqref{eq:CIM}, but $\lambda_j$ and $J_j$ are now exchanged.
The term with the largest coupling which has to be renormalized is now
$\hat{\tilde{H}}_0 = -\lambda_j\hat{\mu}_{j-1}^x\hat{\mu}_{j}^z\hat{\mu}_{j+1}^x$.
Applying to it the same analysis of the first case, we see that the renormalization procedure eliminates the site $j$
in the dual representation, and generates the term
\begin{equation}
 -\widetilde{J}_{j+1}\hat{\mu}_{j-2}^y\hat{\mu}_{j+2}^y\quad{\rm with}\quad \widetilde{J}_j=\dfrac{J_{j-1}J_{j+1}}{\lambda_{j-1}}\,.
  \label{eq:RG_dual}
\end{equation}

In the limit of many RG steps, and after applying the central limit theorem, it is possible to see that 
the disordered CIM is equivalent to a system with couplings 
\begin{eqnarray}
  \widetilde\lambda & = & \exp \big[2l \big( \, \overline{\log \lambda} - \overline{\log J} \, \big) \big] \nonumber\\
  \widetilde J & = & \exp \big[ 2l \big( \, \overline{\log J} - \overline{\log \lambda} \big) \big] \, ,
  \label{eq:RG_params}
\end{eqnarray}
where $l$ is the number of consecutive 
renormalized sites, which in principle can be different for different sites of the renormalized model {(all the details of the calculation are in Appendix~\ref{app:RG})}. 
We can distinguish three cases: \\
\indent I: $\; \overline{\log \lambda}>\overline{\log J}$. In this case, $\widetilde\lambda$ 
is larger than $\widetilde J$ exponentially in the number of iterations of the renormalization step. 
In the limit of infinite iterations, $\widetilde J$ is vanishingly small with respect to $\widetilde\lambda$: 
only the AF terms survive. Therefore, the RG flows to an AF condition and the system is antiferromagnetic. \\
\indent II: $\; \overline{\log \lambda}<\overline{\log J}$. Similarly to case I, 
in the limit of infinite iterations, $\widetilde \lambda$ is vanishingly small with respect 
to $\widetilde J$: only the three-body (cluster) terms survive. 
Looking at the problem in the dual representation, we see that the dual system has only the AF term 
and then the RG in this representation flows to an antiferromagnetic condition. 
Going back to the original representation, we see that the system flows to the SO phase. \\
\indent III: $\; \overline{\log \lambda}=\overline{\log J}$. Here, the RG flows to a uniform system 
with $\widetilde\lambda=\widetilde J = 1$: the low-energy behavior of the model is equivalent 
to a uniform model at the critical point between the SO and the AF phase. 

Therefore, also in the disordered model, we can predict a transition between AF and SO phase,
occurring for 
\begin{equation} \label{dis_trans:eqn}
  \overline{\log \lambda} = \overline{\log J} \,.
\end{equation}
For $\overline{\log \lambda}>\overline{\log J}$ the model is AF, while for 
$\overline{\log \lambda}<\overline{\log J}$ it displays SO.
These results are very similar to those found in the transverse field Ising model~\cite{Fisher_1995}.

\subsection{Numerical analysis of the two phases} \label{CIM_num:sec}

Just to fix the ideas, let us now analyze the case in which there is no disorder on $J$ ($J_j = 1, \; \forall j$),
and each $\lambda_j$ is taken from a uniform distribution among 0 and some ${\lambda}_{\rm max}$.
In this case, the condition in Eq.~\eqref{dis_trans:eqn} implies that the transition point is located at
\begin{equation} \label{lamb_x:eqn}
  \lambda_{{\rm max},c}^{(\infty)} = {\rm e} \approx 2.718\ldots \,,
\end{equation}
${\rm e}$ being the Neper number. The superscript ``$(\infty)$'' in Eq.~\eqref{lamb_x:eqn} denotes that this is the critical point in the limit $L\rightarrow\infty$.
By exploiting the fact that the CIM is exactly solvable, we can explore the behavior of the long-range string and 
AF correlator for considerably long system sizes, after averaging over an ensemble of several disorder realizations. 

Here, we recall that, fixing the realization of the disorder, a Jordan-Wigner transformation is able to map 
Eq.~\eqref{eq:CIM} into a free-fermion diagonal form (irrespective of the presence or absence of translation invariance):
\begin{equation} \label{diag_gamma:eqn}
  \hat{H}=\sum_{\mu=1}^L\epsilon_\mu \big(2 \hat \gamma^\dagger_\mu \hat \gamma_\mu - 1 \big)\,,
\end{equation}
where the single quasi-particle operators $\hat \gamma_\mu$ are defined in terms of the local fermionic
operators $\hat c_j$ according to
\begin{equation} \label{gamma_gamma:eqn}
  \hat \gamma_\mu = \sum_{j=1}^L\left(U_{j\mu}^{\,*} \hat c_j + V_{j\mu}^{\,*} \hat c^\dagger_j \right)\,,
\end{equation}
and $U_{j\mu}^{\,*}$, $V_{j\mu}^{\,*}$ are the coefficients of the $2L\times 2L$ unitary matrix which
diagonalizes the appropriate $2L\times 2L$ Hermitian matrices forming the Hamiltonian 
(see for instance Ref.~\onlinecite{Russomanno_JSTAT13} for more details on this method). 
The GS is the one which is annihilated by all the $\hat \gamma_\mu$ operators (it has a BCS form). 
Thanks to this property, we can evaluate the AF correlator in Eq.~\eqref{eq:neelCIM}.
Applying Wick's theorem to the BCS Gaussian state, we can write the AF correlator as a Toeplitz determinant:
\begin{equation} \label{corr_ev:eqn}
  (\mathcal{S}_{[1/2],l}^\alpha)^2=
  \frac{1}{4} \;\left|\begin{array}{ccc}G_{k,k+1}&\cdots&G_{k,k+l}\\
      \vdots& &\vdots\\
      G_{k+l-1,k+l}&\cdots&G_{k+l-1,k+l}\end{array}\right| \,,
\end{equation}
where, for each disorder realization, we have defined the two point fermionic correlators
on the GS corresponding to that realization:
\begin{equation}
  G_{j,m} = \big\langle (\opcdag{j}-\opc{j})(\opcdag{m}+\opc{m}) \big\rangle \,.
\end{equation}
Inverting Eq.~\eqref{gamma_gamma:eqn} and using the fact that the GS is annihilated by all the $\hat \gamma_\mu$, 
we can evaluate this correlator as
\begin{equation}
  G_{j,m}=\sum_\mu(V_{j\mu}^{\,*}-U_{j\mu})(U_{m\mu}^{\,*}+V_{m\mu})\,.
\end{equation}
The SO parameter in Eq.~\eqref{eq:stringCIM} is evaluated by applying the duality transformation~\eqref{duality0:eqn}, 
which maps it onto an AF correlator of the form in Eq.~\eqref{corr_ev:eqn}, and the Hamiltonian onto 
another Hamiltonian of the same form. Finally, the results obtained through these formulas 
are averaged over $N_{\rm av}$ realizations of disorder.

\begin{figure}[!t]
  \includegraphics[width=0.9\columnwidth]{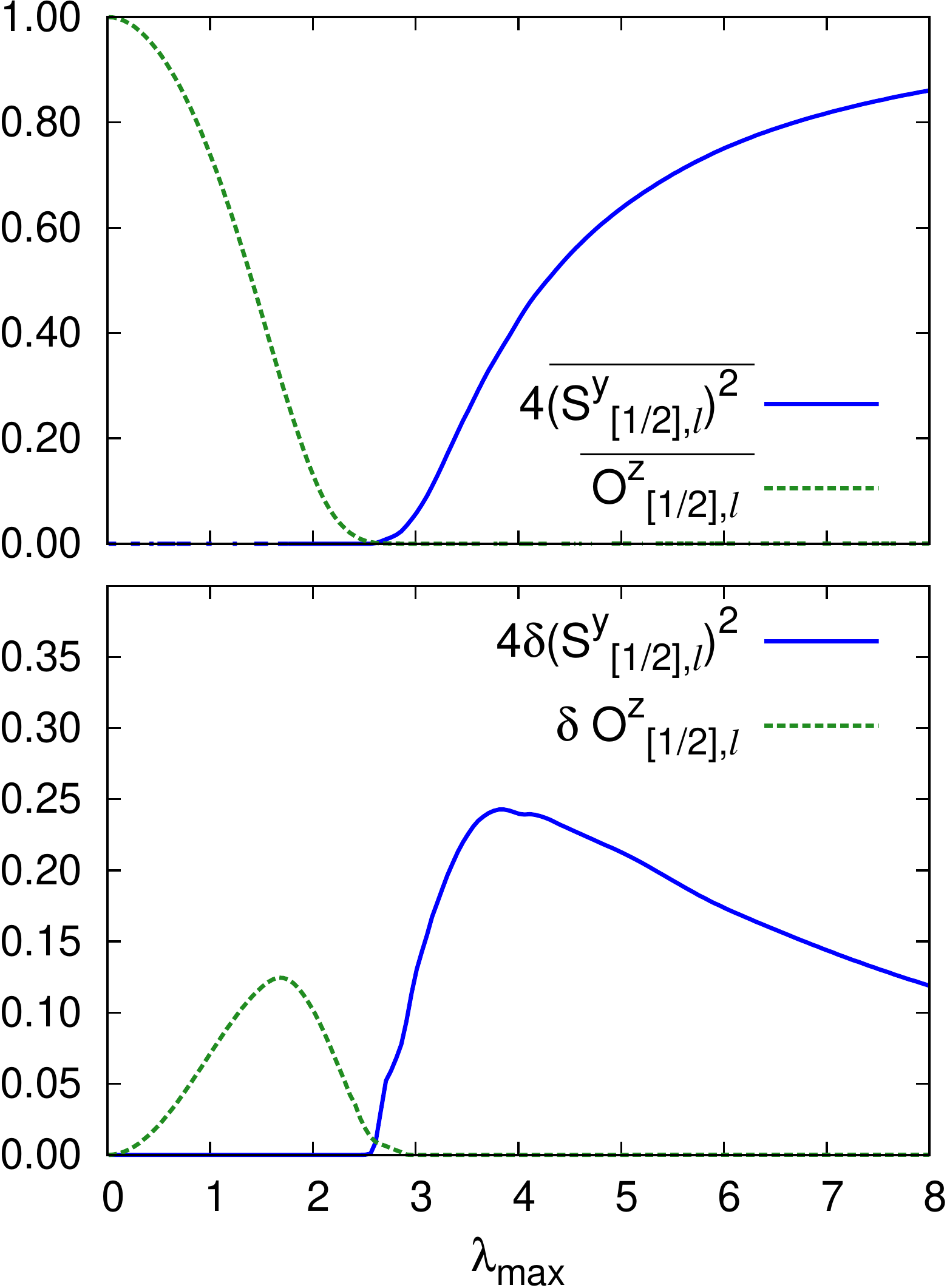}
  \caption{Upper panel: AF (blue) and SO (green) parameters versus $\lambda_{\rm max}$ in the disordered CIM.
    The AF order has been approximated by the finite-range correlator $\overline{(\mathcal{S}_{[1/2],l}^y)^2}$, 
    while the SO by $\overline{\mathcal{O}_{[1/2],l}^z}$ (see Sec.~\ref{CIM_mod:sec}). 
    Lower panel: Fluctuations over the disorder of AF and SO correlators versus $\lambda_{\rm max}$ 
    [see Eq.~\eqref{flucco:eqn}]. 
    We evaluated the correlators between site $l_0=60$ and site $L-60$ (that is $l=L-120$). 
    Data have been obtained for systems with $L=500$ sites, and after averaging 
    over $N_{\rm av}=200$ disorder realizations.}
\label{corr_dis:fig}
\end{figure}

The outcomes of our computations for a given finite size are reported in Fig~\ref{corr_dis:fig}. 
On the upper panel we plot $\overline{\mathcal O^z_{[1/2],l}}$ and $\overline{(\mathcal{S}^y_{[1/2],l})^2}$. 
In order to avoid unwanted boundary effects, we evaluate the correlators between sites 
that are far away from the chain ends (see the caption for details).
These quantities approximate the order parameters of Eqs.~\eqref{eq:neelCIM} and ~\eqref{eq:stringCIM}, 
since in numerical simulations we can consider large, but yet finite system sizes. 
We see in the upper panel of Fig~\ref{corr_dis:fig} that, when $\overline{{\mathcal O}^z_{[1/2],l}}$ vanishes, 
$\overline{(\mathcal{S}^y_{[1/2],l})^2}$ appears with a crossover: 
this is an indication that there is a transition from a string-ordered to a $y$-antiferromagnetic phase 
in the thermodynamic limit, even in the presence of disorder. 

We can also analyze fluctuations over the disorder of the two order parameters. 
As before, we calculate finite-range correlators. Focusing on the AF order, 
we can define the fluctuation as
\begin{equation} \label{flucco:eqn}
  \delta (\mathcal{S}^y_{[1/2],l})^2 = \Big[ \overline{\langle \hat{\sigma}_k^y \hat{\sigma}_{k+l}^y \rangle^2}-\overline{\langle \hat{\sigma}_k^y \hat{\sigma}_{k+l}^y \rangle}^2 \Big]^{1/2} \, ,
\end{equation}
where the expectation value has to be intended over the GS of any specific disorder realization. 
The definition for the SO fluctuation ($\delta\mathcal O^z_{[1/2],l}$) is analogous, 
after replacing the correlator of Eq.~\eqref{eq:neelCIM} with that of Eq.~\eqref{eq:stringCIM}. 
The results are shown in the lower panel of Fig.~\ref{corr_dis:fig}. We can see that fluctuation is different from zero 
only when the corresponding order parameter is nonvanishing (compare with the upper panel): 
the finite-size crossover appears also in the behavior of fluctuations.

Until now, we have considered signatures of the transition in finite-length correlators. 
As we can observe in Fig.~\ref{corr_dis:fig}, finite-size effects are evident: for a given system size, 
we actually see a crossover, and there is a region where both the finite-range order parameters are different from zero. 
Moreover, if we identify the transition with the point where the curves of the two finite-range order parameters cross, 
we get a result that is different from the theoretical prediction~\eqref{lamb_x:eqn}. 
In order to properly infer the behavior in the thermodynamic limit $L,l\to\infty$, 
we perform a finite-size scaling analysis. 

\begin{figure}[!t]
  \includegraphics[width=\columnwidth]{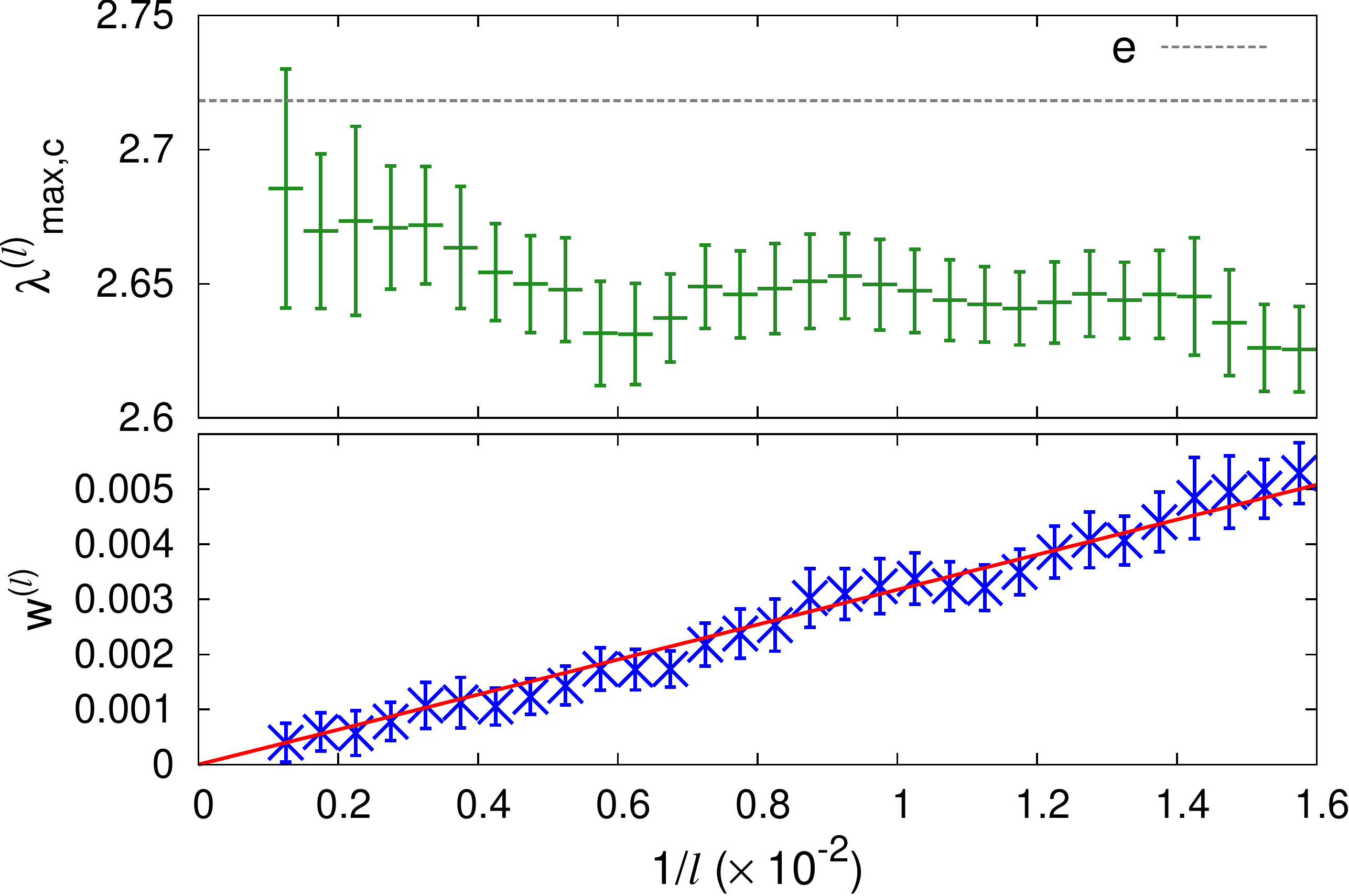}
  \caption{Upper panel: Location of the finite-size crossing point $\lambda_{{\rm max},c}^{(l)}$
    between AF and SO parameters versus $1/l$: for $1/l\to 0$ the crossing point tends to the theoretical
    transition value of Eq.~\eqref{lamb_x:eqn}, consistently with the error bars resulting from disorder-averaging+coarse-graining fluctuations (see the main text). 
    Lower panel: the height $w^{(l)}$ of the crossing point vanishes for $1/l\to 0$ as confirmed
    by the straight-line fit (red line -- see the main text). (Numerical parameters: $l_0 = 60,\,N_{\rm av}=1000,\,\delta(1/l)=5\cdot 10^{-4},\,L=1100$).
}
  \label{trappa:fig}
\end{figure}

In order to reduce the effect of the fluctuations induced by the noise, we need to perform a coarse graininig in $1/l$ of disorder-averaged correlators. 
More precisely, we proceed in the following way. We fix the value of $L$ and, in order to avoid finite-size boundary effects, we fix an appropriate $l_0$ and consider the disorder-averaged correlator between the site $l_0$ and the site $l$ with $l_0+l$ varying between 0 and $L - 2l_0$. Then, we coarse grain this correlator: we consider the interval in which {the quantity} $1/l$ varies, divide this interval in windows of width $\delta(1/l)$ and perform the average of the correlator over each window. For each of the resulting values we evaluate the uncertainty as 
the maximum over the corresponding window of the disorder fluctuation of the correlator: applying the central limit theorem, this fluctuation is given by the value in Eq.~\eqref{flucco:eqn}
divided by $\sqrt{N_{\rm av}}$. 
We label each of the windows over which we coarse grain with its central value $1/l$: for each value of $1/l$ we locate the approximate transition point  as the value of $\lambda_{\rm max}$ where the two coarse-grained correlators cross; we denote the crossing point as $\lambda_{{\rm max},c}^{(l)}$. In the upper panel of Fig.~\ref{trappa:fig} we show $\lambda_{{\rm max},c}^{(l)}$ versus $1/l$: taking into account the error bars, we see a behaviour consistent with a convergence towards
the theoretical value of Eq.~\eqref{lamb_x:eqn} when $1/l\to0$. 
The error bars are evaluated in the following way: each time we add or subtract the uncertainties discussed above to the coarse-grained disorder-averaged 
correlators and then we  take the half-dispersion of the 4 resulting estimates of the crossing. 

The lower panel of Fig.~\ref{trappa:fig} shows that the height $w^{(l)}$ of the crossing point tends to zero when $1/l\to0$ (the error bars are evaluated with the same method used for $\lambda_{{\rm max},c}^{(l)}$). 
We have checked this fact by fitting with a straight line $a\cdot(1/l)$, and found 
 {$(a=317\pm 4)\times10^{-5}$} (red line in Fig.~\ref{trappa:fig}).
This means that, in the thermodynamic limit, $w^{(\infty)}=0$, and thus there is no region where both the order parameters 
are {non-}vanishing: when one vanishes the other appears, as appropriate for a QPT. 

\subsection{Edge modes}

For a clean system, the QPT in the spin chain maps to a topological transition in the fermionic representation. 
In the case of periodic boundary conditions, the AF phase corresponds to winding number one in the fermionic picture, 
while the SO phase corresponds to winding number two~\cite{Smacchia_2011}.~\cite{Note_winding}
Taking OBC, the topological nature of the system appears through the existence 
of zero-energy boundary modes~\cite{kitaev2001unpaired, Alicea_2012}: diagonalizing the fermionic Hamiltonian, 
some vanishing $\epsilon_\mu$ appear in Eq.~\eqref{diag_gamma:eqn}. 
For each phase, there is a fixed number of zero-energy modes, and their amplitudes $U_{j}$ and $V_j$ 
[see Eq.~\eqref{gamma_gamma:eqn}] are localized on the edges of the system. 
The AF phase displays one zero-energy mode, while the SO phase has two zero-energy modes (edge modes
in uniform fermionic Hamiltonians very similar to Eq.~\eqref{eq:CIM_fermion} have been studied in Refs.~\onlinecite{prb85,prb88}).

\begin{figure}[!t]
  \includegraphics[width=\columnwidth]{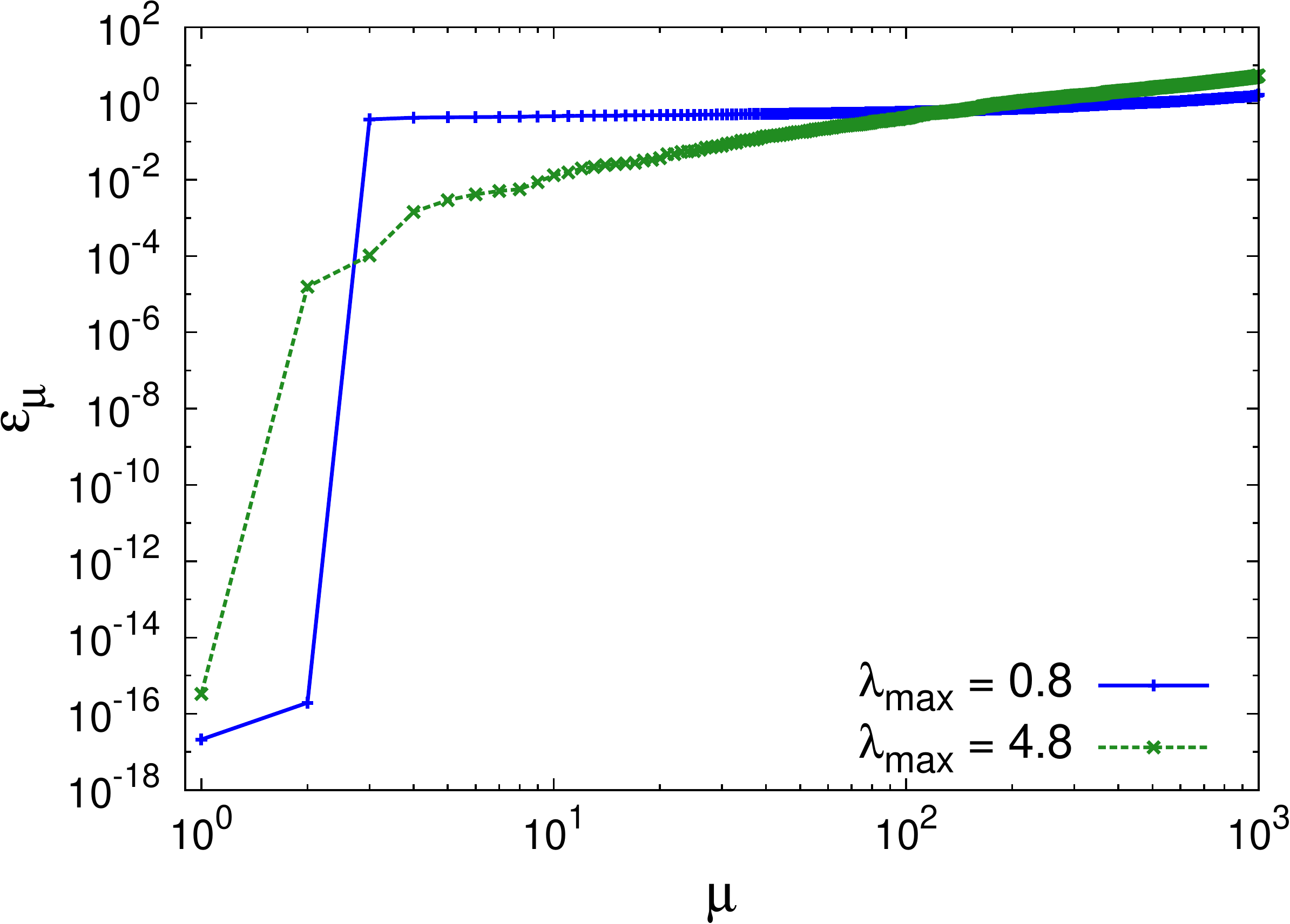}
  \caption{Single quasi-particle energy eigenvalues $\epsilon_\mu$ in increasing order for two values of $\lambda_{\rm max}$. 
    For $\lambda_{\rm max}=0.8$, the system is in the SO phase and displays two zero-energy boundary modes 
    (they appear as two energy eigenvalues with energy many orders of magnitude smaller than the others).
    For $\lambda_{\rm max}=4.8$, the system is in the AF phase and shows one boundary mode. 
    Here we considered $L=1000$ and took one single realization of disorder.}
    \label{dis_incre:fig}
\end{figure}

Even in the presence of disorder, we numerically observe the persistence in the spectrum of zero-energy modes
(two modes in the SO phase, and one mode in the AF phase). 
Two examples of this fact are reported in Fig.~\ref{dis_incre:fig}. 
Here, we choose a specific disorder realization and show the single quasi-particle spectrum $\epsilon_\mu$ 
for a case where the system shows SO ($\lambda_{\rm max}=0.8$) and a case where it is AF ($\lambda_{\rm max}=4.8$). 
In the first situation there are two levels with energy many orders of magnitude smaller than the others; 
in the second one there is a single level with this property: these levels correspond to the boundary modes discussed above 
(the energy is not exactly zero, due to the numerical round-off errors). 
We have verified that the same structure of the spectrum appears for any disorder realization. 

These edge modes are topologically protected, since they only depend on global properties of the system: 
they cannot be destroyed by local perturbations (like disorder) if the perturbation is weak enough. 
That is why, if we add disorder, two edge modes and the associated SO persists for $\lambda_{\rm max}$ small, 
and one edge modes and AF order persist for $\lambda_{\rm max}$ large. 
For $\lambda_{\rm max}$ around the transition, the disorder is strong enough to move the transition point.

\subsection{{Thermodynamic singularities}}
\begin{figure}[!t]
  \includegraphics[width=\columnwidth]{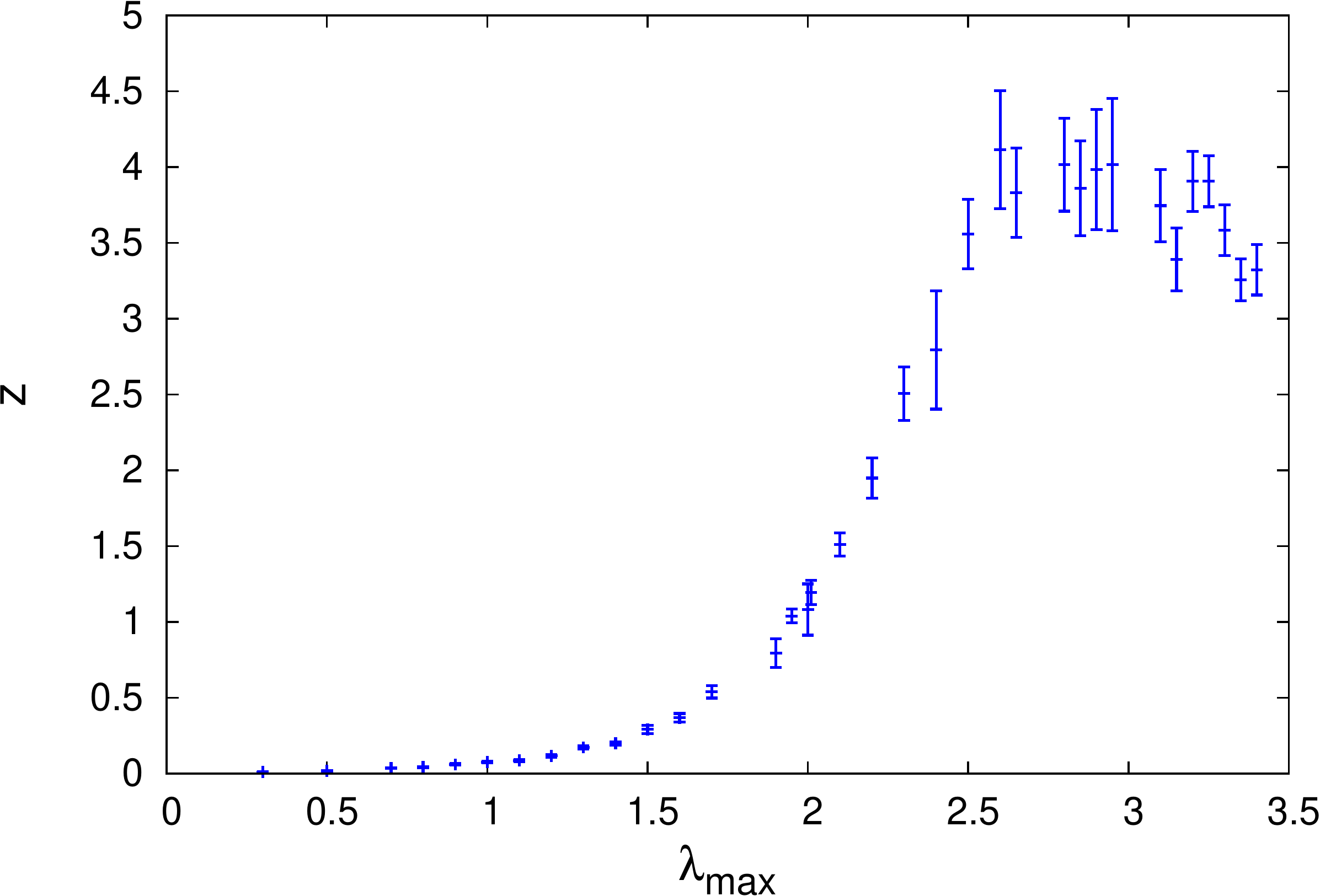}
  \caption{Plot of the dynamic exponent $z$ obtained from the fit of the logarithmic level spacing distribution $P(\log\Delta)$ versus $\lambda_{\rm max}$.  For $\lambda_{\rm max}$ above a threshold $\lambda_{\rm max}^{c\,s}\simeq2$ it becomes larger than 1 giving rise to a singular behaviour. (Numerical parameters: $N_{\rm av}=2000,\,L=400$, periodic boundary conditions).
}
  \label{zazza:fig}
\end{figure}

\begin{figure}[!t]
  \includegraphics[width=\columnwidth]{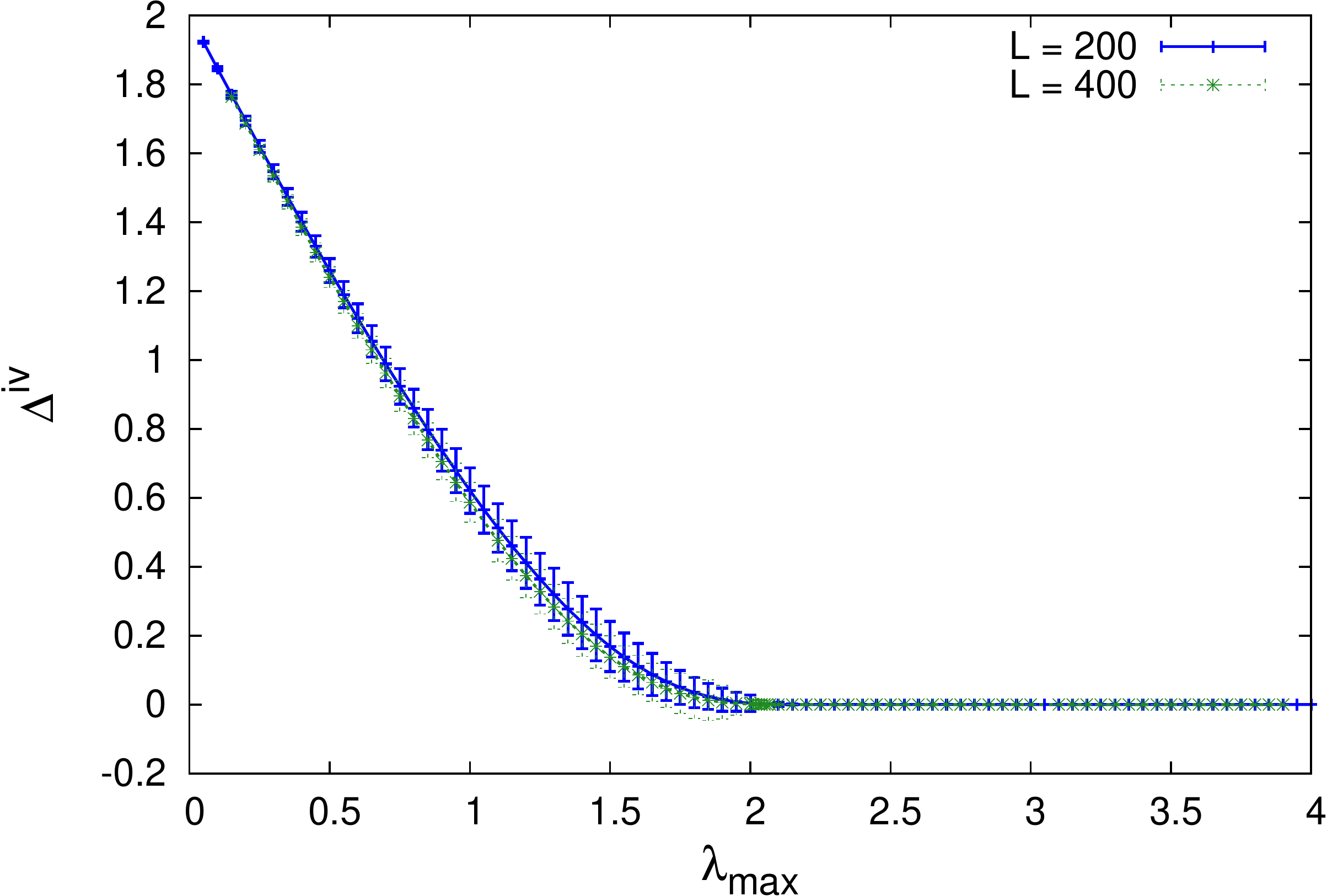}
  \caption{Plot of the inverse average of the gap versus $\lambda_{\rm max}$. It vanishes for $\lambda_{\rm max} > \lambda_{\rm max}^{c\,s}=2$: this parameter range corresponds to a singular region ($z>1$). (Numerical parameters: $N_{\rm av}=2000$, periodic boundary conditions).
}
  \label{Inverse_gap:fig}
\end{figure}

{Disordered systems can display phases where the thermodynamic quantities show singularities in the limit of vanishing temperature~\cite{Griffiths,Lajko_PRB,Boechat}. This can be seen from the behaviour of the distribution of the energy gap $\Delta$ of the Hamiltonian.
If for small $\Delta$ the logarithmic energy gap distribution behaves as $P(\log\Delta)\sim\Delta^{1/z}$ ($z$ is the so-called dynamic exponent), it
is easy to show that the excitation energy over the ground state behaves as $E_{\rm ex}(T)\sim T^{1+1/z}$ at low temperatures. Therefore, the low-temperature specific heat
behaves as $C\sim T^{1/z}$ and its derivative shows a divergence in the limit of $T\to 0$ when $z>1$. The ranges of parameters where this happens are called singular regions~\cite{Lajko_PRB}; in order to find them we numerically consider the properties of $P(\log\Delta)$ and we check that it behaves as a power law for small $\Delta$ (see some examples in Fig.~\ref{distribuzze:fig}).
\begin{figure}[!t]
  \includegraphics[width=\columnwidth]{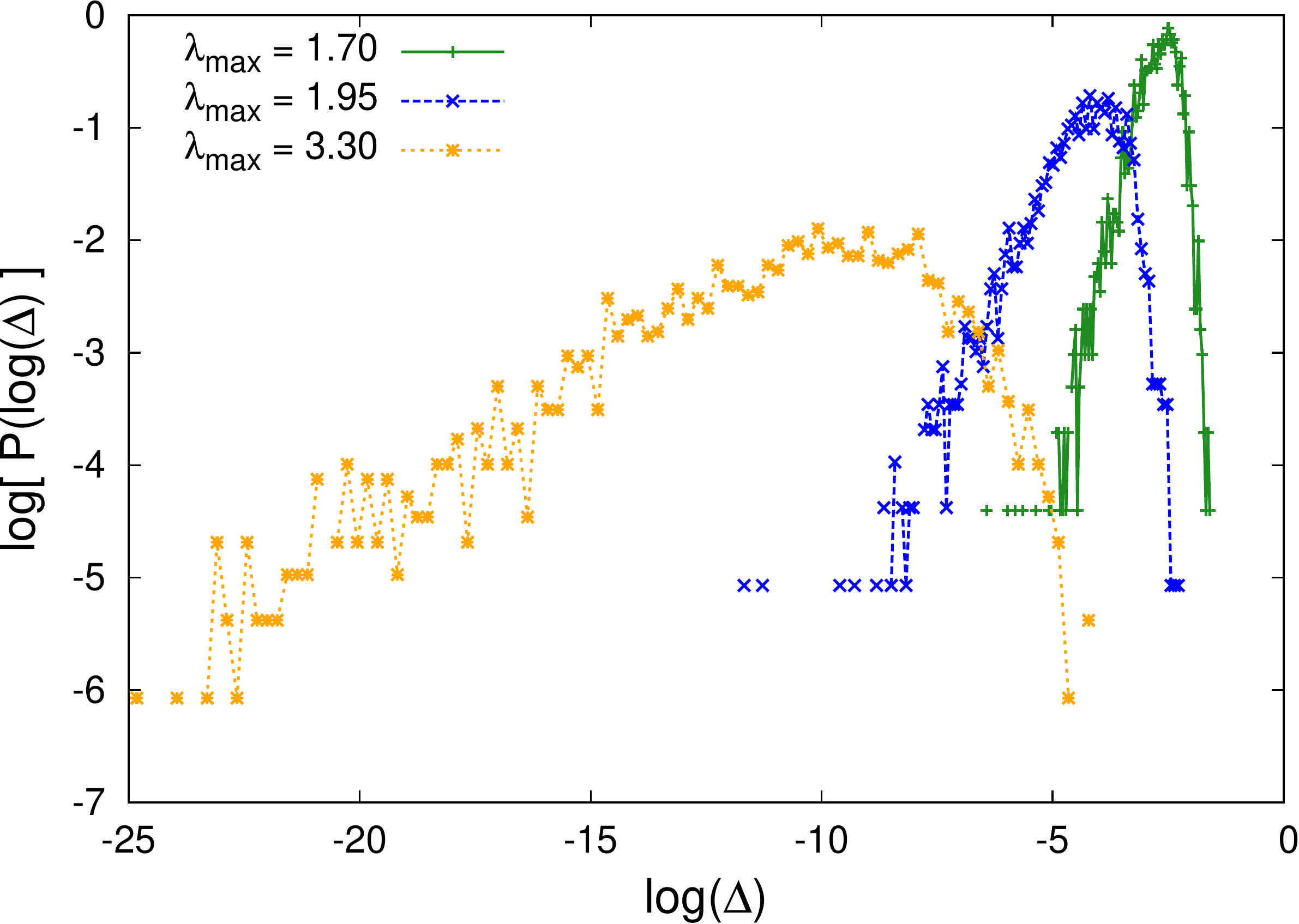}
  \caption{Logarithmic level spacing distributions: in the bi-logarithmic plots the power-law at small $\Delta$ appears as a linear behaviour. We use periodic boundary conditions in order to avoid the boundary zero-energy modes. (Numerical parameters: $N_{\rm av}=2000,\,L=400$).
}
  \label{distribuzze:fig}
\end{figure}
Applying a linear fit to the plots of $\log[P(\log\Delta)]$ vs $\log\Delta$ we are able to estimate the value of the dynamic exponent $z$ which we plot in Fig.~\ref{zazza:fig}. We see that there is a singular region with $z>1$ for $\lambda_{\rm max}$ above a threshold $\lambda_{\rm max}^{c\,s}\simeq2$. In order to have a better estimate of $\lambda_{\rm max}^{c\,s}$, we follow Ref.~\onlinecite{Lajko_PRB} and consider the behaviour of the inverse average of the gap defined as
\begin{equation}
  \Delta^{\rm iv} \equiv\left[\overline{\left(\frac{1}{\Delta}\right)}\right]^{-1}\,,
\end{equation}
where the average is performed over the disorder distribution. This object vanishes whenever the system is in a singular region~\cite{Lajko_PRB} with $z>1$; we show results for our case in Fig.~\ref{Inverse_gap:fig}. We see that $\Delta^{\rm iv}$ vanishes for $\lambda_{\rm max}>\lambda_{\rm max}^{c\,s}=2$, confirming that in this parameter range our system is singular. Moreover, we can numerically find that $\Delta^{\rm iv}$ vanishes as a power law when $\lambda$ approaches the transition point $\lambda_{\rm max}^{c\,s}$ from below: we have $\Delta^{\rm iv}\sim(\lambda_{\rm max}^{c\,s}-\lambda_{\rm max})^{\mu_{\Delta}}$ with $\mu_{\Delta}=2.06\pm0.01$.}

{In conclusion, we see a transition to a singular regime which occurs at a critical value $\lambda_{\rm max}^{c\,s}$ different from the critical point $\lambda_{\max,\,c}$ separating the SO and the AF phase. While for $\lambda_{\rm max}>\lambda_{{\rm max},\,c}$ the system is AF and singular, we have a non-singular  SO phase ($z<1$) for $\lambda_{\rm max}<\lambda_{\rm max}^{c\,s}$ and  there is a singular SO phase ($z>1$) for $\lambda_{\rm max}^{c\,s}<\lambda_{\rm max}<\lambda_{\max,\,c}$. This behaviour is strictly reminiscent the disordered $S=1$ antiferromagnetic Heisenberg chain~\cite{Lajko_PRB} where there is a gapped Haldane phase ($z<1$) and a singular Haldane phase ($z>1$), both showing SO. We emphasize that the singularity structure of the phases
is strictly related to the specific form of the disorder: for instance, taking $\lambda$ uniform and $J_j$ uniformly distributed between 0 and some $J_{\rm max}$, we would have seen a non-singular AF phase together with a singular AF phase, while the SO phase would have been fully non-singular.}
\section{Haldane to N\'eel phase transition in the disordered Spin-1 XXZ model} \label{XXZ:sec}

We now switch to study the zero-temperature properties of the disordered spin-1 XXZ Hamiltonian 
of Eq.~\eqref{eq:XXZ}. 
Since this model is non integrable, in order to find the GS of a given finite-size system,
we resort to a variational search on the class of MPS~\cite{Schollwock2011}.
In our simulations, we analyze chains of up to $L=240$ sites and
choose a maximum bond link $D_{\rm max}=400$. We set $J$ as the reference energy scale{, and consider $J=1$ in the following}.
As we have done in the CIM, to characterize the two phases we focus on the finite-system AF correlation 
function $\big( {\mathcal S}^z_{[1],l} \big)^2$ and on the $x$-axis string operator ${\mathcal O}^x_{[1],l}$
[the corresponding order parameters are defined, in the thermodynamic limit, by Eqs.~\eqref{eq:SO-1} and~\eqref{eq:AF-1}].
Since we have $\big( {\mathcal S}^z_{[1],l} \big)^2\geq0$ and $\mathcal{O}^x_{[1],l}\leq0$ in the XXZ chain, in the following we will consider the absolute value of the string parameter in order to deal with positive quantities.

\begin{figure}[!t]
  \includegraphics[width=\columnwidth]{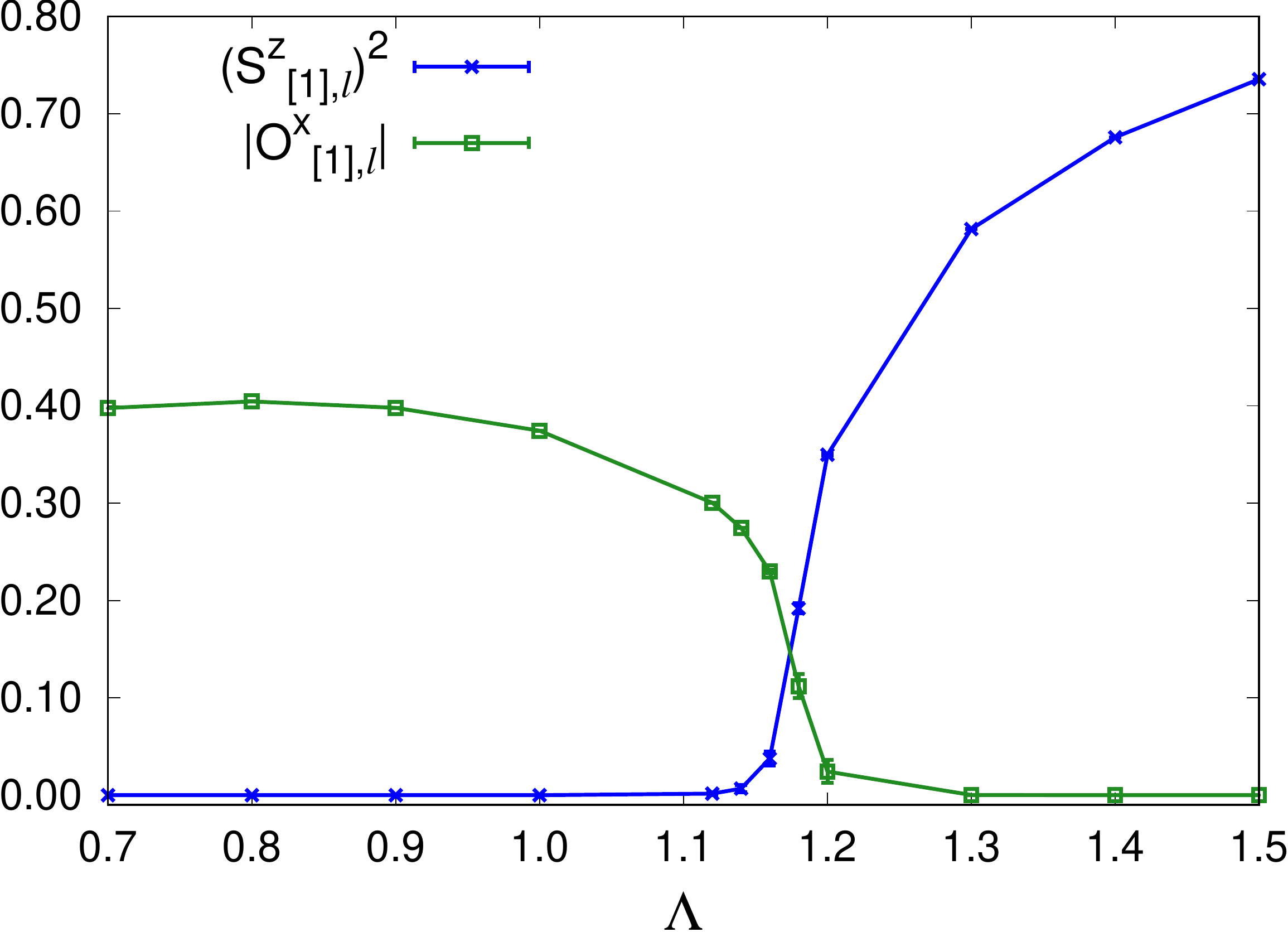}
  \caption{The AF (blue) and SO (green) parameters for the clean XXZ model 
    (${\Lambda}_j={\Lambda}, \; \forall j$), as a function of the anisotropy ${\Lambda}$. 
    As we did for the CIM, the AF order has been approximated by the finite-size 
    correlator ${({\mathcal S}^z_{[1],l})}^2$, while the SO by $|{\mathcal O}^x_{[1],l}|$ (see Sec.~\ref{Heis:XXZ:subsec}). 
    Here we simulated a chain of $L=240$ sites and evaluated the correlators
    between site $k=24$ and site $k+l=120$ (that is $l=96$),
    as also explained in Fig.~\ref{corr_dis:fig} for the CIM.
    We locate the transition point at the crossing of the two curves, that is ${\Lambda}^{(L)}_{c}=1.17 \pm 0.01$. 
    {We estimate the uncertainty over ${\Lambda}^{(L)}_{c}$ as} half of the discretization 
    of ${\Lambda}$ in proximity of the crossing point.}
\label{fig:phasetransition240sitesclean}
\end{figure}

Before analyzing in detail the phase transition in the presence of disorder, let us briefly discuss the clean XXZ model.
The situation is summarized in Fig.~\ref{fig:phasetransition240sitesclean}, which shows
the behavior of the bulk expectation value of the staggered correlator along $z$ (blue data set),
and of the absolute value of the string correlator along $x$ (green data set).
We see that, at a critical value ${\Lambda}^{(L)}_c$ of the anisotropy term, the SO vanishes
and the staggered order starts to take a finite value: as explained before, this is an indication
of the occurrence of the Haldane-N\'eel phase transition.
The position of the critical point in the thermodynamic limit, ${\Lambda}^{(\infty)}_c$, can be inferred from the finite-size scaling of the crossing point between the two curves; from our simulations at finite $L$, we estimate ${\Lambda}^{(L)}_c=1.17\pm0.01$, which is in agreement with the value ${\Lambda}^{(\infty)}_c \simeq 1.186\dots$ found in the thermodynamic limit~\cite{DegliEsposti_2004,Ueda_2008}.

Based on our knowledge on the clean XXZ model, we now focus on the Haldane-N\'eel phase transition 
in the disordered scenario. 
Since MPS simulations are computationally more demanding and can only afford systems 
with a comparatively small length, we adopted a procedure  slightly different from the one used
for the CIM, in order to estimate the SO correlator [Eq.~\eqref{eq:SO-1}] and the AF correlator [Eq.~\eqref{eq:AF-1}]
from the bulk expectation values of the corresponding finite-size correlators.
The two methods coincide in the thermodynamic limit, but the one described here {is more appropriate for the smaller
value of system size and number of disorder realizations which we can obtain with DMRG in the XXZ chain because it enables to
minimize the uncertainty in the averages.}

\begin{figure}[t]
  \includegraphics[width=\columnwidth]{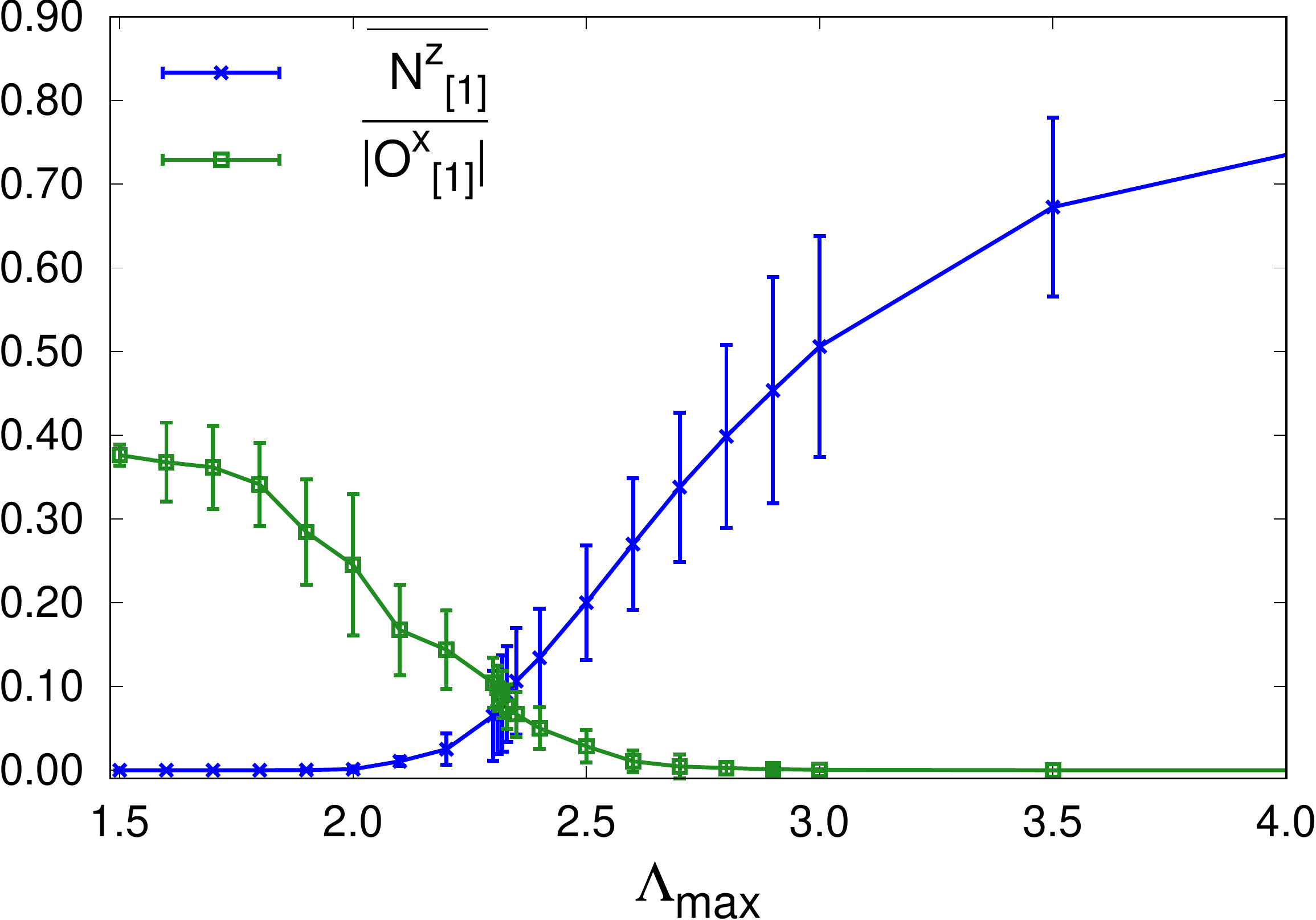}
  \caption{AF (blue) and SO (green points) correlators averaged over space and over
    $N_{\rm av}=40$ realizations of disorder [Eqs.~\eqref{eq:spatialaverageofa} and~\eqref{eq:disorderaverageofa}], 
    as a function of ${\Lambda}_{\rm max}$. 
    Here we use $L=120$, ${\Lambda}_{\rm min}=0$, and locate the transition 
    at the intersection of the two curves: ${\Lambda^{(L)}_{\rm max,c}=2.32 \pm 0.01}$. 
}
  \label{fig:phasetransition120sites}
\end{figure}

Namely, we are interested in computing the bulk expectation values of a given two-point observable, 
of the form $\hat{\mathcal{A}}_{k,k+l}$. 
We recall that both the AF correlator, $\big( \mathcal{S}^\alpha_{[1],l} \big)^2$, 
and the SO correlator, $\mathcal{O}^\alpha_{[1],l}$, can be seen as expectation values of observables 
which live on a given number of sites in between the $k$-th and the $(k+l)$-th site.
After fixing the system size and the disorder realization, we first compute a space average 
over different lengths $l$ of the correlator~\cite{Note1}, in order to average out space fluctuations.
We discard the sites that are close 
to the two chain ends, thus disregarding boundary effects 
(see Appendix~\ref{sec:appendixdetailsonthekinks} for details).
Then, we repeat the simulation by varying the configuration of the disorder, and eventually perform 
a second average of such obtained space averages, over different disorder realizations.
{The obtained correlators have an uncertainty  (denoted by error bars in Fig.~\ref{fig:phasetransition120sites}) which
is estimated by computing the variance of the space averaged correlators for the different realizations of disorder, as detailed in Appendix~\ref{sec:appendixdetailsonthekinks}}.

We first present our numerical analysis for ${\Lambda}_j$ uniformly distributed in the interval $[0,{\Lambda}_{\rm max}]$, 
for all $j$. In this case, we expect the system to undergo the Haldane-N\'eel phase transition 
as ${\Lambda}_{\rm max}$ is varied across some critical value.
In our simulations we see that, in the N\'eel phase, the AF pattern is affected by the presence 
of kinks (domain walls, where the AF pattern is reversed), which hide the presence of long-range AF order~\eqref{eq:AF-1}. 
As detailed in Appendix~\ref{sec:appendixdetailsonthekinks}, the presence of such kinks is a numerical artifact 
due to the non-perfect convergence of the MPS algorithm. Thus, instead of computing the staggered correlator 
as in Eq.~\eqref{eq:AF-1}, we can get rid of the kinks and reveal the presence of AF long-range order 
by computing the bulk average [Eq.~\eqref{eq:spatialaverageofa}] of the N\'eel correlator, which is defined as
\begin{equation}
\mathcal{N}^z_{[1],l} = \big| \langle \hat S^z_k \hat S^z_{k+l} \rangle \big| \, .
\label{eq:neelcorrelator}
\end{equation}
Notice that $\big( \mathcal{S}^z_{[1],l} \big)^2$ coincides with the N\'eel correlator 
in Eq.~\eqref{eq:neelcorrelator} in the case of perfect AF order (no kinks), 
but differently from the staggered correlator, $\mathcal{N}^z_{[1],l}$ is insensitive 
to such numerical artifacts, because of the presence of the absolute value in Eq.~\eqref{eq:neelcorrelator}. 
Thus, we characterize Haldane and N\'eel phases in the disordered chain by looking respectively
at the SO along $x$ [see Eq.~\eqref{eq:SO-1}] and the N\'eel correlator 
in the limit of $l \to \infty$: $\mathcal{N}^z_{[1]} = \lim_{l\to\infty} \mathcal{N}^z_{[1],l}$.

\begin{figure}[t]
  \includegraphics[width=\columnwidth]{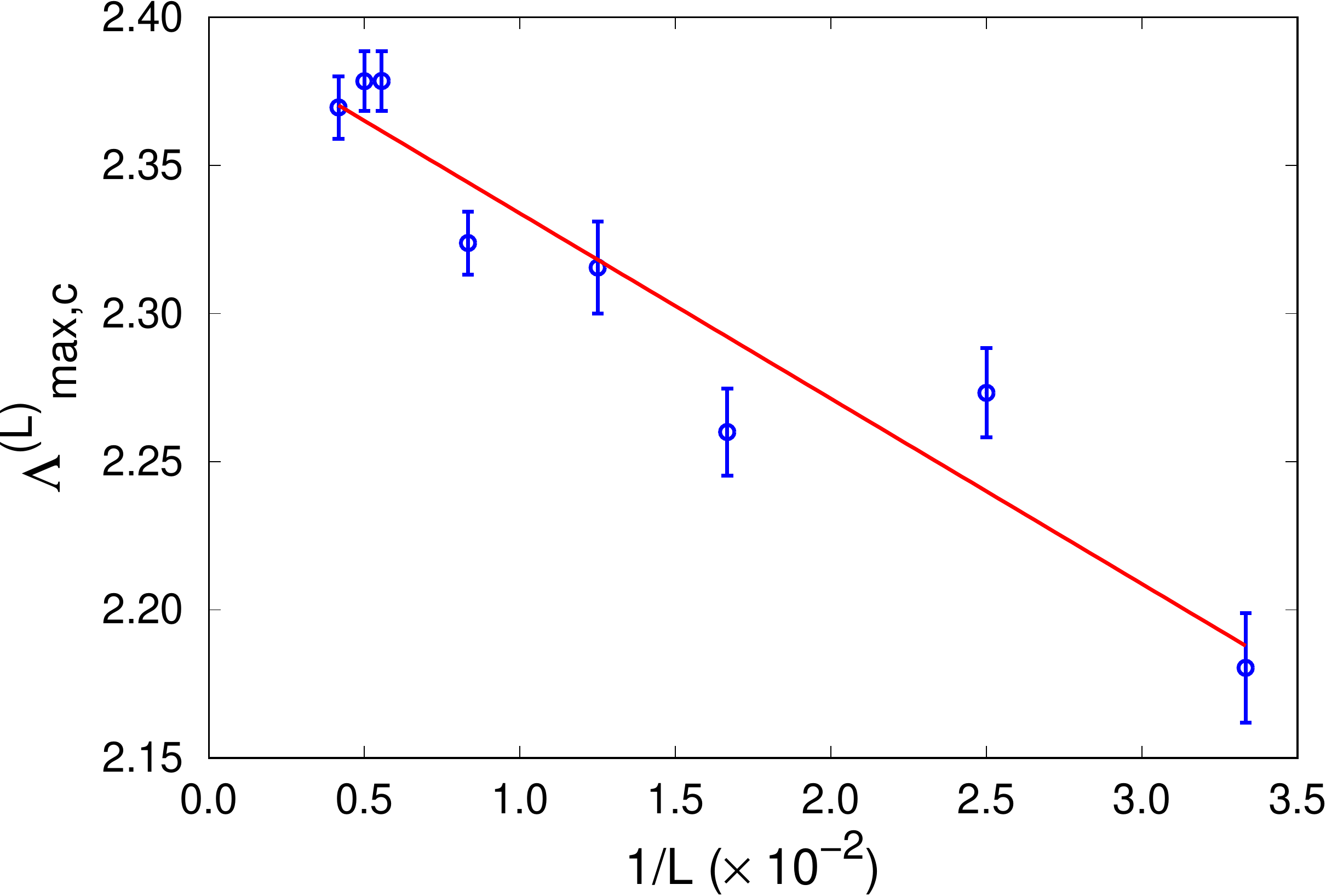}
  \caption{Finite-size scaling for the critical point ${\Lambda}_{{\rm max},c}^{(L)}$ with ${\Lambda}_{\rm min}=0$. 
    The data are shown as a function of $1/L$. Blue points correspond to the numerical data of ${\Lambda}_{{\rm max},c}^{(L)}$, 
    (the uncertainty is estimated as in Fig.~\ref{fig:phasetransition240sitesclean}).
    The red solid line is the fit with the function $f(L)=a/L+{\Lambda}_{{\rm max},c}^{(\infty)}$, 
    treating $a$ and ${\Lambda}_{{\rm max},c}^{(\infty)}$ as fit parameters. 
    Here, ${\Lambda}_{{\rm max},c}^{(\infty)}$ is the critical point in the thermodynamic limit, 
    which we estimate as ${\Lambda_{{\rm max},c}^{(\infty)}=2.40\pm0.01}$.}
  \label{fig:finitesizescalingcriticalpoint}
\end{figure}

In Fig.~\ref{fig:phasetransition120sites}, we show the result of simulations with $L=120$ and ${\Lambda}_{\rm min}=0$,
after averaging over space and over disorder. 
Blue points correspond to the N\'eel order, $\overline{\mathcal{N}^z_{[1]}}$, 
whereas green points are the SO data, $\overline{|\mathcal{O}^x_{[1]}|}$. 
We observe that the N\'eel order is zero for sufficiently small ${\Lambda}_{\rm max}$, and starts to increase 
around a given value of ${\Lambda}_{\rm max}$. Conversely, the SO along $x$ is nonzero for small ${\Lambda}_{\rm max}$, 
and goes to zero as ${\Lambda}_{\rm max}$ is increased. This behavior is analogous to the one for the
Haldane-N\'eel phase transition in the clean XXZ model (Fig.~\ref{fig:phasetransition240sitesclean}). 
Furthermore, we can estimate a critical point ${\Lambda}^{(L)}_{\rm max,c}$ which is shifted 
with respect to the clean value: for the simulation in Fig.~\ref{fig:phasetransition120sites}, 
we find ${\Lambda_{\rm max,c}^{(L)}=2.32\pm0.02}$. For a given finite size the transition behaves as a crossover, 
exactly as it occurs for the CIM (upper panel of Fig.~\ref{corr_dis:fig}) {and we estimate the finite-$L$ approximation of the critical
point as the abscissa of the crossing point of the two curves of the correlators. The error bars of the crossing points are estimated as follows. We consider the plot of the correlators (for instance Fig.~\ref{fig:phasetransition120sites}) and, being interested in the fluctuations of the intersection of the disorder
averages, we divide the error bars by $\sqrt{N_{\rm av}}$, according to the central limit theorem. Later, we proceed in a way similar to Fig.~\ref{trappa:fig}: each time we add or subtract these fluctuations to the disorder-averaged 
correlators and then we  take  the half-dispersion of the 4 resulting values of the crossing.  This gives the uncertainty of the crossing point.}

As in the CIM case, in order to extrapolate the value of the critical point 
in the thermodynamic limit, we need to perform a finite-size scaling analysis and
repeat the same simulations as in Fig.~\ref{fig:phasetransition120sites} for different values of $L$. 
The result is shown in Fig.~\ref{fig:finitesizescalingcriticalpoint}:
blue points correspond to the estimated values of ${\Lambda}_{\rm max,c}^{(L)}$, and uncertainties
are computed as explained above. 
To find the thermodynamic value of the critical point, we show the data as a function of $1/L$ 
and perform a best fit with the function $f(L)=a/L+{\Lambda}_{{\rm max},c}^{(\infty)}$. 
From the result in Fig.~\ref{fig:finitesizescalingcriticalpoint}, 
we extrapolate the critical value in the asymptotic $L\to\infty$ limit: ${\Lambda_{{\rm max},c}^{(\infty)}=2.40\pm0.01}$.

\begin{figure}[t]
  \includegraphics[width=\columnwidth]{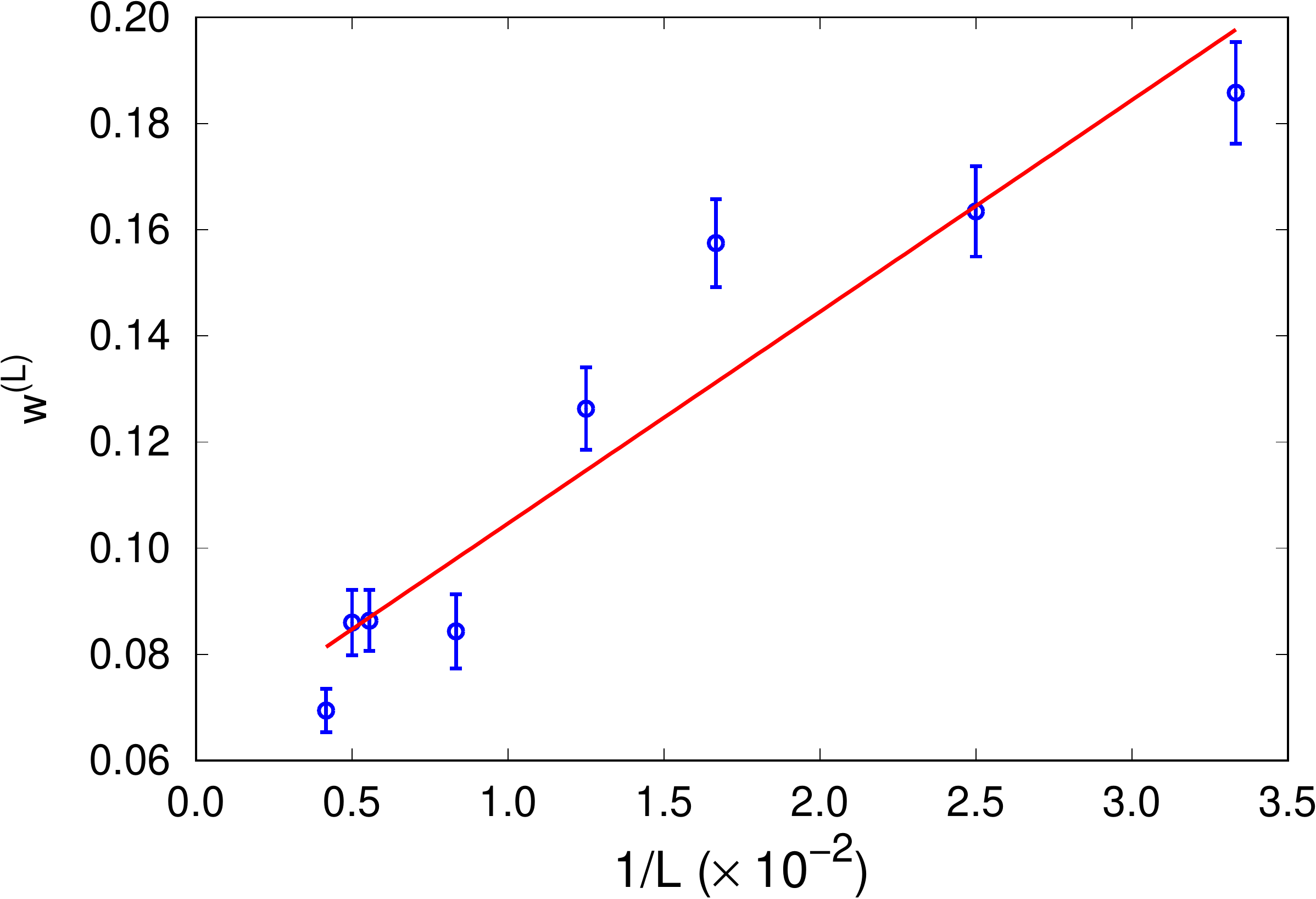}
  \caption{Finite-size scaling for the height of the crossing point $w^{(L)}$ with ${\Lambda}_{\rm min}=0$. 
    Data (symbols) are shown as a function of $1/L$, and their uncertainties are computed 
    from those of the disorder-averaged correlators. 
    We see that the height of the crossing point decreases in the thermodynamic limit, 
    meaning that the transition becomes sharper and sharper as $L$ is increased.}
  \label{fig:finitesizescalingheigth}
\end{figure}

We can also define the height of the crossing point 
$w^{(L)}$, as in Sec.~\ref{CIM_num:sec}. For each value of $L$, the uncertainty on the value of $w^{(L)}$ is computed from that of the disorder-averaged correlators. We show the result in Fig.~\ref{fig:finitesizescalingheigth}, 
from which we see that $w^{(L)}$ decreases as $L$ is increased. From the fit, we estimate ${w^{(\infty)}=0.06\pm0.01}$. 
This is an indication of the fact that the phase transition becomes sharper and sharper as $L$ is increased.

\begin{figure}[!t]
  \includegraphics[width=\columnwidth]{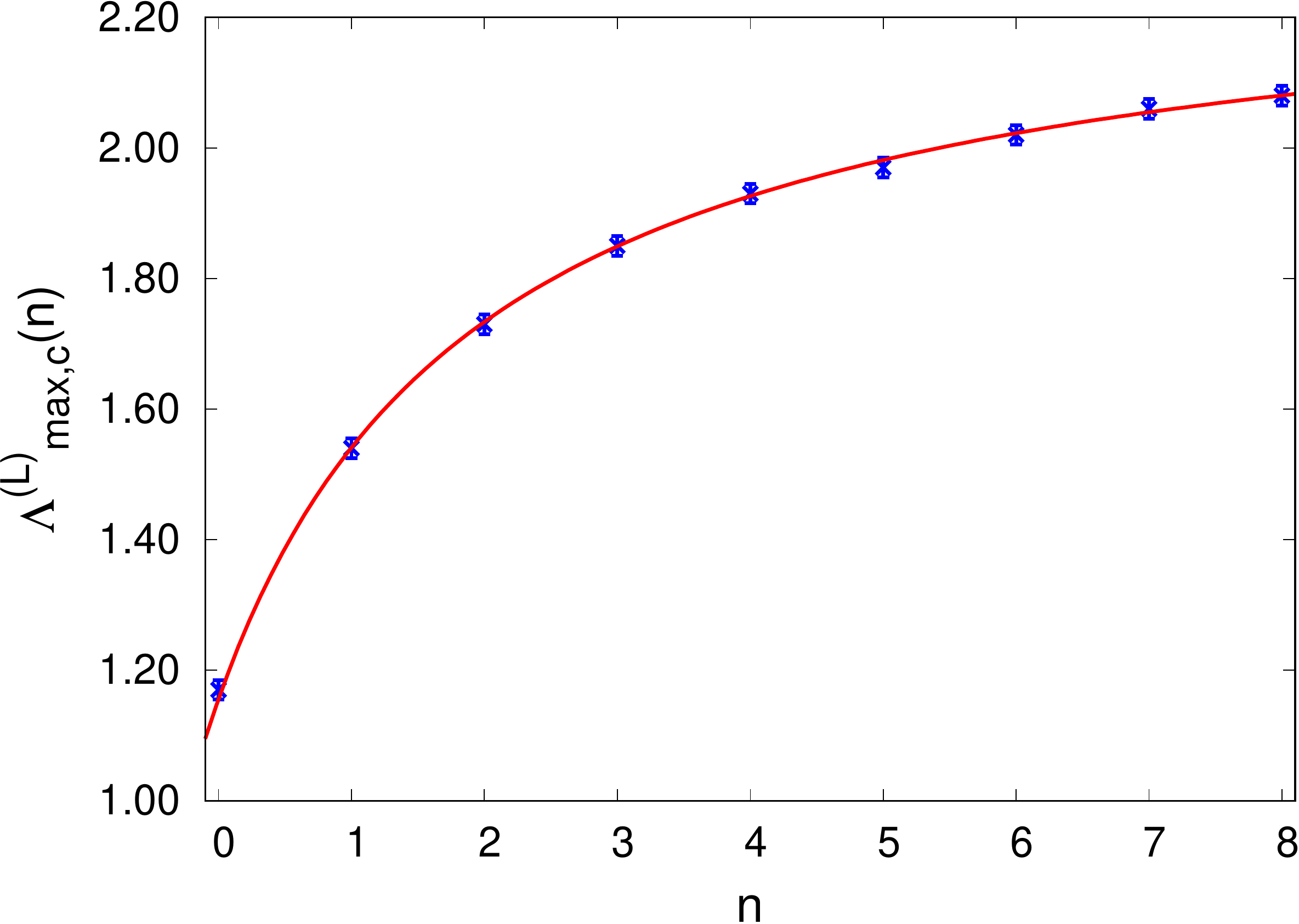}
  \caption{Position of the critical point for the Haldane-N\'eel phase transition, 
    for ${\Lambda}_j\in[{\Lambda}_{\rm max}/(n+1),{\Lambda}_{\rm max}]$, for all $j$, where $n$ is an integer number. 
    Here we use $L=120$ and average over $N_{\rm av}=40$ realizations of disorder. 
    The value of the critical point ${\Lambda}_{{\rm max},c}^{(L)}(n)$, for a given $n$, is estimated as explained for the data 
    in Fig.~\ref{fig:phasetransition240sitesclean}. 
    The red curve corresponds to the fit with the function in Eq.~\eqref{eq:fittingdataxxz}.}
  \label{fig:nscalingcriticalpoint}
\end{figure}

So far, we have discussed the Haldane-N\'eel phase transition for ${\Lambda}_j$ uniformly distributed 
in the interval $[{\Lambda}_{\rm min},{\Lambda}_{\rm max}]$, for all $j$, using ${\Lambda}_{\rm min}=0$. 
In order to see how the position of the critical point is affected by the choice of ${\Lambda}_{\rm min}$ 
and ${\Lambda}_{\rm max}$, we simulate the disordered model of Eq.~\eqref{eq:XXZ} 
varying ${\Lambda}_{\rm max}$ and using ${\Lambda}_{\rm min}={\Lambda}_{\rm max}/(n+1)$, where $n$ is a positive integer number.
Our results for $L=120$ are shown in Fig.~\ref{fig:nscalingcriticalpoint}. 
For each value of $n$, we estimate the position of the critical point ${\Lambda}_{{\rm max},c}^{(L)}(n)$ 
as explained for the data in Fig.~\ref{fig:phasetransition240sitesclean}. 
If we define the algebraic average of $\{{\Lambda}_j\}$ in the chain, i.e.
\begin{equation}
  {\bar\Lambda}=\frac{{\Lambda}_{\rm min}+{\Lambda}_{\rm max}}{2}=\frac{{\Lambda}_{\rm max}}{2}\,\frac{n+2}{n+1} \, ,
  \label{eq:meanvaluexxzchain}
\end{equation}
we find that the Haldane-N\'eel phase transition in the disordered chain occurs when ${\Lambda}_{\rm max}$ is such that 
the mean value of $\{{\Lambda}_j\}$ in Eq.~\eqref{eq:meanvaluexxzchain} equals the critical value ${\Lambda}_c$ 
of the clean chain, i.e., inverting Eq.~\eqref{eq:meanvaluexxzchain} and showing explicitly the dependence on $L$:
\begin{equation}
  {\Lambda}_{{\rm max},c}^{(L)}(n)=2{\Lambda}^{(L)}_c\,\frac{n+1}{n+2} \, .
  \label{eq:fittingdataxxz}
\end{equation}
As is evident from Fig.~\ref{fig:nscalingcriticalpoint}, the position of the critical point agrees with 
the scaling given by Eq.~\eqref{eq:fittingdataxxz}, where we use ${\Lambda}^{(L)}_c$ as fit parameter. 
From the fit, we estimate ${\Lambda^{(L)}_c=1.157\pm0.002}$, which is not in disagreement with the clean value ${\Lambda}^{(L)}_c \simeq 1.17$ 
found in Fig.~\ref{fig:phasetransition240sitesclean} for $L=240$. 
We ascribe the slight discrepancy of the two estimates to finite-size effects (we use $L=120$ for the data in Fig.~\ref{fig:nscalingcriticalpoint}).

{Differently from the CIM, the study of the thermodynamic singularities in the case of the XXZ model
  requires a much larger computational effort, due to the increased numerical complexity. This is left as an open issue for a future work.}

\section{Conclusions and perspectives} \label{conco:sec}
%
In conclusion, in this work we have studied the existence of the non-local string order in disordered spin chains.
We have focused on two models, the spin-$1/2$ cluster Ising chain and the spin-$1$ XXZ Heisenberg chain, which are
well known to show a transition from antiferromagnetism to string order in the clean case (the first model is moreover interesting for applications
in quantum information~\cite{Smacchia_2011,Son_2011}). We have discovered that this
transition persists in both cases if disorder is added. 

In the disordered cluster Ising model we have found a transition from antiferromagnetism to string order by numerically studying the order parameters in
the ground state; we did this using the Jordan-Wigner mapping on an integrable free-fermion model. We have seen also that the transition manifests
in the fermionic representation as a change in the number of 0-energy edge modes. In this model we have found analytically the position of the
transition point using the strongest coupling renormalization group: this analytical prediction is fully confirmed by the finite-size-scaling
on our numerical results. {Moreover, studying the thermodynamic singularity at vanishing temperature, we have found a transition between a non-singular and a singular behaviour when the strength of the disorder is increased. We have seen that this transition point is different from the one separating the SO and the AF phase.}

In the disordered spin-1 XXZ Heisenberg chain we have studied the order parameters in the ground state by means of the DMRG technique: we have found a transition between an
antiferromagnetic and a string-ordered phase and we have determined its position by means of finite-size scaling. This model is very interesting
because its string-ordered phase is adiabatically connected to the celebrated Haldane phase~\cite{Haldane_1983} and it can be experimentally studied thanks
to the new cold-atom techniques~\cite{senko_PRX}.

Perspectives of future work include first of all the application of the spin-1 strongest coupling renormalization group~\cite{Boechat, Boechat1} to the XXZ Heisenberg chain, in order to
analytically predict the phase transition point that we find here numerically. Here
we have only addressed the properties of the ground state: it will be interesting 
to consider the properties of all the spectrum, in connection with many-body 
localization (MBL)~\cite{Basko_Ann06, Oganesyan_PRA75} of interacting non-integrable system. MBL systems can show topological order 
in a large fraction of the excited energy eigenstates~\cite{Chitta_JSTAT13}: it would be interesting to see
if also our non-integrable spin-1 disordered XXZ model shows MBL and if string and antiferromagnetic order persist in excited states. 
A possibility to study these phenomena is applying to the Hamiltonian a quantum quench and look at the dynamics 
of the string correlator. In a clean spin-1 XXZ Heisenberg model the string thermalizes~\cite{Rossini_PRB14,Strinati_PRB16}, 
only the ground state being ordered, but in disordered systems the situation 
could be much different thanks to the MBL. 
\vspace{0.3cm}


\section*{Acknowledgements}
We are grateful to A.~Hamma and L.~Mazza for fruitful discussions. We thank E.~G.~Dalla~Torre for useful comments
on the manuscript. We acknowledge the CINECA award under the ISCRA initiative, for the availability of high performance computing resources and support. {M.~C.~S. acknowledges partial support from the Israel Science Foundation, grants No.~231/14 and~1452/14}. A.~R. acknowledges financial support from EU through project QUIC, from ``Progetti interni - Scuola Normale Superiore'' and from his parents. R.~F. kindly acknowledges support from EU through project QUIC (Grant Agreement No.~641122), the National
Research Foundation of Singapore (CRP - QSYNC) and the Oxford Martin School.


\appendix

\section{Details on the renormalization group for the CIM}
\label{app:RG}

In the renormalization procedure of the CIM described in Sec.~\ref{strongest:sec}, we can distinguish
two different types of RG steps: {\em i)} the largest coupling is a $J_j$ 
or {\em ii)} the largest coupling is a $\lambda_j$.

\emph{Case i)}:
If we assume that the largest coupling is $J_j$, it is not difficult to see that the local Hamiltonian $\hat{H}_0$
of Eq.~\eqref{eq:Hloc_RG} has four degenerate ground states:
\begin{eqnarray}
  \ket{1}&=&\ket{(+)_{j-1}\,\uparrow_j\,(+)_{j+1}}\,,\nonumber\\
  \ket{2}&=&\ket{(-)_{j-1}\,\uparrow_j\,(-)_{j+1}}\,,\nonumber\\
  \ket{3}&=&\ket{(-)_{j-1}\,\downarrow_j\,(+)_{j+1}}\,,\nonumber\\
  \ket{4}&=&\ket{(+)_{j-1}\,\downarrow_j\,(-)_{j+1}}\,,
\end{eqnarray}
where $\ket{\uparrow_l}$, $\ket{\downarrow_l}$ are the eigenstates of $\hat{\sigma}_l^z$ and $\ket{(+)_l}$, $\ket{(-)_l}$ are the eigenstates of $\hat{\sigma}_l^x$.

Applying the degenerate perturbation theory to such four ground states, we get the following corrections at first
order in the perturbation $\hat{V}$ of Eq.~\eqref{eq:V_RG}:
\begin{widetext}
\begin{eqnarray}
  \ket{\psi_{g1}} & = & \ket{(+)_{j-1}\,\uparrow_j\,(+)_{j+1}}
                       +\frac{i}{2J_j}\left[\lambda_{j-1}\hat{\sigma}_{j-2}^y\ket{(-)_{j-1}\,\uparrow_j\,(+)_{j+1}}
                       +\lambda_{j+1}\hat{\sigma}_{j+2}^y\ket{(+)_{j-1}\,\uparrow_j\,(-)_{j+1}}\right] \,, \\
  \ket{\psi_{g2}} & = & \ket{(-)_{j-1}\,\uparrow_j\,(-)_{j+1}}
                       -\frac{i}{2J_j}\left[\lambda_{j-1}\hat{\sigma}_{j-2}^y\ket{(+)_{j-1}\,\uparrow_j\,(-)_{j+1}}
                       +\lambda_{j+1}\hat{\sigma}_{j+2}^y\ket{(-)_{j-1}\,\uparrow_j\,(+)_{j+1}}\right] \,, \\
  \ket{\psi_{g3}} & = & \ket{(-)_{j-1}\,\downarrow_j\,(+)_{j+1}}
                       -\frac{i}{2J_j}\left[\lambda_{j-1}\hat{\sigma}_{j-2}^y\ket{(+)_{j-1}\,\uparrow_j\,(+)_{j+1}}
                       -\lambda_{j+1}\hat{\sigma}_{j+2}^y\ket{(-)_{j-1}\,\uparrow_j\,(-)_{j+1}}\right] \,, \\
  \ket{\psi_{g4}} & = & \ket{(+)_{j-1}\,\downarrow_j\,(-)_{j+1}}
                       -\frac{i}{2J_j}\left[-\lambda_{j-1}\hat{\sigma}_{j-2}^y\ket{(-)_{j-1}\,\uparrow_j\,(-)_{j+1}}
                       +\lambda_{j+1}\hat{\sigma}_{j+2}^y\ket{(+)_{j-1}\,\uparrow_j\,(+)_{j+1}}\right] \, .
\end{eqnarray}
To apply degenerate perturbation theory, we construct and diagonalize the matrix $V_{ij}=\bra{i}\hat{V}\ket{\psi_{gj}}$
which is given by:
\begin{equation}
  \mathbb{V}=\left(\begin{array}{cc|cc}
    -\dfrac{\lambda_{j-1}^2+\lambda_{j+1}^2}{2J_j}&\dfrac{\lambda_{j-1}\lambda_{j+1}}{J_j}\hat{\sigma}_{j-2}^y\hat{\sigma}_{j+2}^y&0&0\\
    \dfrac{\lambda_{j-1}\lambda_{j+1}}{J_j}\hat{\sigma}_{j-2}^y\hat{\sigma}_{j+2}^y & -\dfrac{\lambda_{j-1}^2+\lambda_{j+1}^2}{2J_j}&0&0\\ \hline 
    0&0&-\dfrac{\lambda_{j-1}^2+\lambda_{j+1}^2}{2J_j}&\dfrac{\lambda_{j-1}\lambda_{j+1}}{J_j}\hat{\sigma}_{j-2}^y\hat{\sigma}_{j+2}^y\\
    0&0&\dfrac{\lambda_{j-1}\lambda_{j+1}}{J_j}\hat{\sigma}_{j-2}^y\hat{\sigma}_{j+2}^y&-\dfrac{\lambda_{j-1}^2+\lambda_{j+1}^2}{2J_j}
  \end{array}\right) \,.
\end{equation}
\end{widetext}
This matrix can be written as
\begin{equation}
  V_{ij}=\sum_{n}^{\textrm{excited states}}\frac{\bra{i}\hat{V}\ket{n}\bra{n}\hat{V}\ket{j}}{E_{\rm GS}-E_n}\,,
\end{equation}
where $E_{\rm GS}$ is the energy of the degenerate ground states $\ket{i}$, $\ket{j}$.
Diagonalizing this matrix, one finds the perturbed ground-state eigenenergies at second order in $\lambda$: 
{ $-\dfrac{\lambda_{j-1}^2+\lambda_{j+1}^2}{2J_j}\pm\dfrac{\lambda_{j-1}\lambda_{j+1}}{J_j}\hat{\sigma}_{j-2}^y\hat{\sigma}_{j+2}^y$.}
We select one of these four eigenstates discarding the others. We arbitrarily choose one of the two states with eigenvalue 
{ $-\dfrac{\lambda_{j-1}^2+\lambda_{j+1}^2}{2J_j}-\dfrac{\lambda_{j-1}\lambda_{j+1}}{J_j}\hat{\sigma}_{j-2}^y\hat{\sigma}_{j+2}^y$.}
Through the renormalization we have indeed eliminated the site $j$ and generated a new coupling
\begin{equation}
  -\widetilde{\lambda}_j\hat{\sigma}_{j-2}^y\hat{\sigma}_{j+2}^y\quad{\rm with}\quad\widetilde{\lambda}_j\simeq \frac{\lambda_{j-1}\lambda_{j+1}}{J_j}\,.
\end{equation}
These operators $\hat{\sigma}_{j-2/j+2}^y$ are in principle different from the unrenormalized ones: they coincide with them up to terms quartic in $\lambda/J$.

\emph{Case ii)}:
If the largest coupling is one of the $\lambda_j$, we can reduce to the first case by applying to the Hamiltonian
Eq.~\eqref{eq:CIM} the duality transformation~\cite{Smacchia_2011}:
\begin{equation} \label{duality:eqn}
  \hat{\mu}_j^x = \prod_{k=1}^j\hat{\sigma}_k^z \,, \qquad
  \hat{\mu}_j^z = \hat{\sigma}_j^x\hat{\sigma}_{j+1}^x\,.
\end{equation}
We find the Hamiltonian in the dual representation as
\begin{equation} \label{Ham:duale:eqn}
  \hat{\tilde{H}}= - \sum_j \Big[ J_j \hat{\mu}_{j-1}^y\hat{\mu}_j^y + \lambda_j\hat{\mu}_{j-1}^x\hat{\mu}_{j}^z\hat{\mu}_{j+1}^x \Big]\,.
\end{equation}
Indeed we can see that, in the limit $L\to\infty$ we are considering, this Hamiltonian is equal to its dual in Eq.~\eqref{eq:CIM},
with $\lambda_j$ and $J_j$ exchanged. The term with the largest coupling which has to be renormalized
is indeed $\hat{\tilde{H}}_0 = -\lambda_j\hat{\mu}_{j-1}^x\hat{\mu}_{j}^z\hat{\mu}_{j+}^x$.
Applying to it the same analysis of the first case, we see that the renormalization procedure eliminates the site $j$
in the dual representation 
and generates the term in Eq.~\eqref{eq:RG_dual}.

It is now easy to show that, after many RG steps, the couplings are renormalized according to
\begin{eqnarray}
  \widetilde{J}_j&=&\frac{J_{j-2l}J_{j-2l+2}\cdots J_{j+2l-2}J_{j+2l}}{\lambda_{j-2l-1}\lambda_{j-2l+1}\cdots\lambda_{j+2l-5}\lambda_{j+2l-3}} \,,\nonumber\\
  \widetilde{\lambda}_j&=&\frac{\lambda_{j-2l}\lambda_{j-2l+2}\cdots \lambda_{j+2l-2}\lambda_{j+2l}}{J_{j-2l+1}J_{j-2l+3}\cdots J_{j+2l-3}J_{j+2l-1}} \,.
\end{eqnarray}
Applying the central limit theorem we find
\begin{eqnarray}
  \log\widetilde{J}_j & = & 2l( \, \overline{\log J}-\overline{\log \lambda} \,) \\
  && + \sqrt{2l}\left(\sqrt{{\rm Var}[\log J] + {\rm Var}[\log\lambda]}\right) u_J \, ,\nonumber \\
  \log\widetilde{\lambda}_j & = & 2l(\overline{\log \lambda}-\overline{\log J}) \\
  && + \sqrt{2l}\left(\sqrt{{\rm Var}[\log J] + {\rm Var}[\log \lambda]}\right) u_\lambda \, , \nonumber
\end{eqnarray}
where $u_J$ and $u_\lambda$ are normally distributed random variables; the averages $\overline{(\ldots)}$ 
and the variances ${\rm Var}[ \ldots ]$ are performed over the distributions of $J_j$ and $\lambda_j$. 
We thus see that, in the limit of infinite RG steps, Eqs.~\eqref{eq:RG_params} hold.

\section{Details on the numerical analysis of disordered Heisenberg chains}

\label{sec:appendixdetailsonthekinks}
In this appendix, we provide details on the strategy that we adopted in order to compute 
the bulk expectation values of generic two-point observables of the form $\hat{\mathcal{A}}_{k,k+l}$,
for the numerical results that have been obtained with the MPS-based algorithm on the spin-$1$ XXZ Heisenberg chain.
We also comment on the analysis of the presence of domain walls in our simulations. 

\subsection{Bulk expectation values}

To compute the bulk expectation values of $\hat{\mathcal{A}}_{k,k+l}$,
we first fix $L$, ${\Lambda}_{\rm min}$, ${\Lambda}_{\rm max}$, and a given instance of disorder.
For each realization, we numerically compute the space bulk-average by discarding 
a certain number $\Delta L$ of sites that are close to the chain ends.
Moreover we consider distances $l > \Delta L$ such that, provided we are
far from the transition point, they are larger than the system's correlation length~\cite{Note1}.
Near the critical point the correlation length tends to diverge, therefore we have always finite-size effects:
in order to understand the properties of the transition in the thermodynamic limit it is thus very important to perform a finite-size scaling as we do in the main text.
Specifically, if the sites are labeled from $1$ to $L$, 
we choose $k=\Delta L=0.2\,L$ and average the expectation values from $l=l_1=0.3\,L$ 
to $l=l_2=0.6\,L$, i.e.:
\begin{equation}
\mathcal{A}_{{\rm avg},h} =
\frac{1}{l_2-l_1}\sum_{l=l_1}^{l_2} \big\langle \hat{\mathcal{A}}_{\Delta L,\Delta L+l} \big\rangle_h \, ,
\label{eq:spatialaverageofa}
\end{equation}
where the subscript ${}_{{\rm avg},h}$ denotes the space average for the $h$-th realization
of disorder.  Then we repeat the simulation by varying the configuration of the disorder in the chain, 
and perform an average over all the $N_{\rm av}$ realizations:
\begin{equation}
  \overline{\mathcal{A}_{\rm avg}} =
  \frac{1}{N_{\rm av}}\sum_{h=1}^{N_{\rm av}} \mathcal{A}_{{\rm avg},h} \, .
\label{eq:disorderaverageofa}
\end{equation}

\begin{figure*}[t]
  \centering
  \includegraphics[width=0.68\columnwidth]{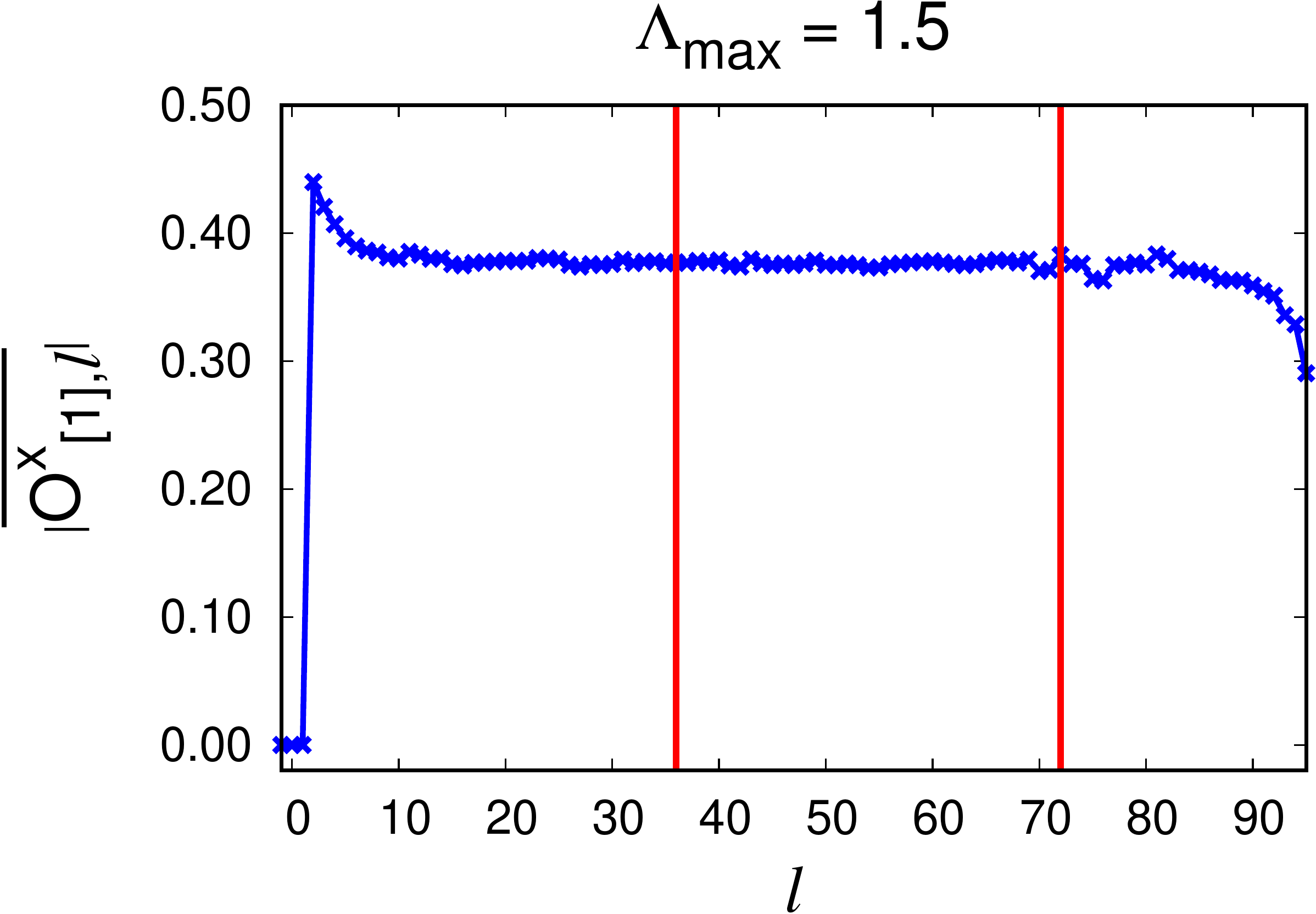}
  \includegraphics[width=0.68\columnwidth]{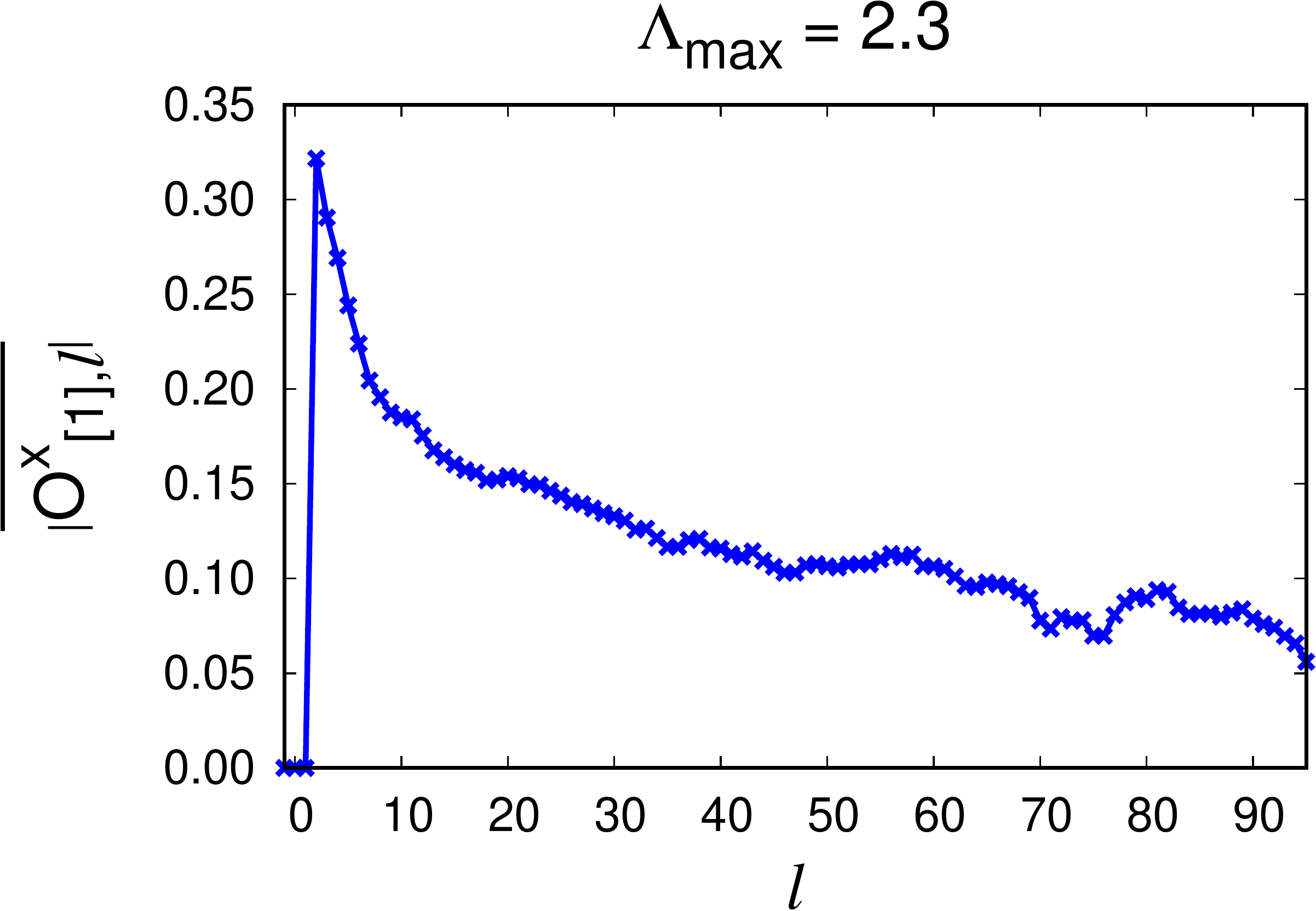}
  \includegraphics[width=0.68\columnwidth]{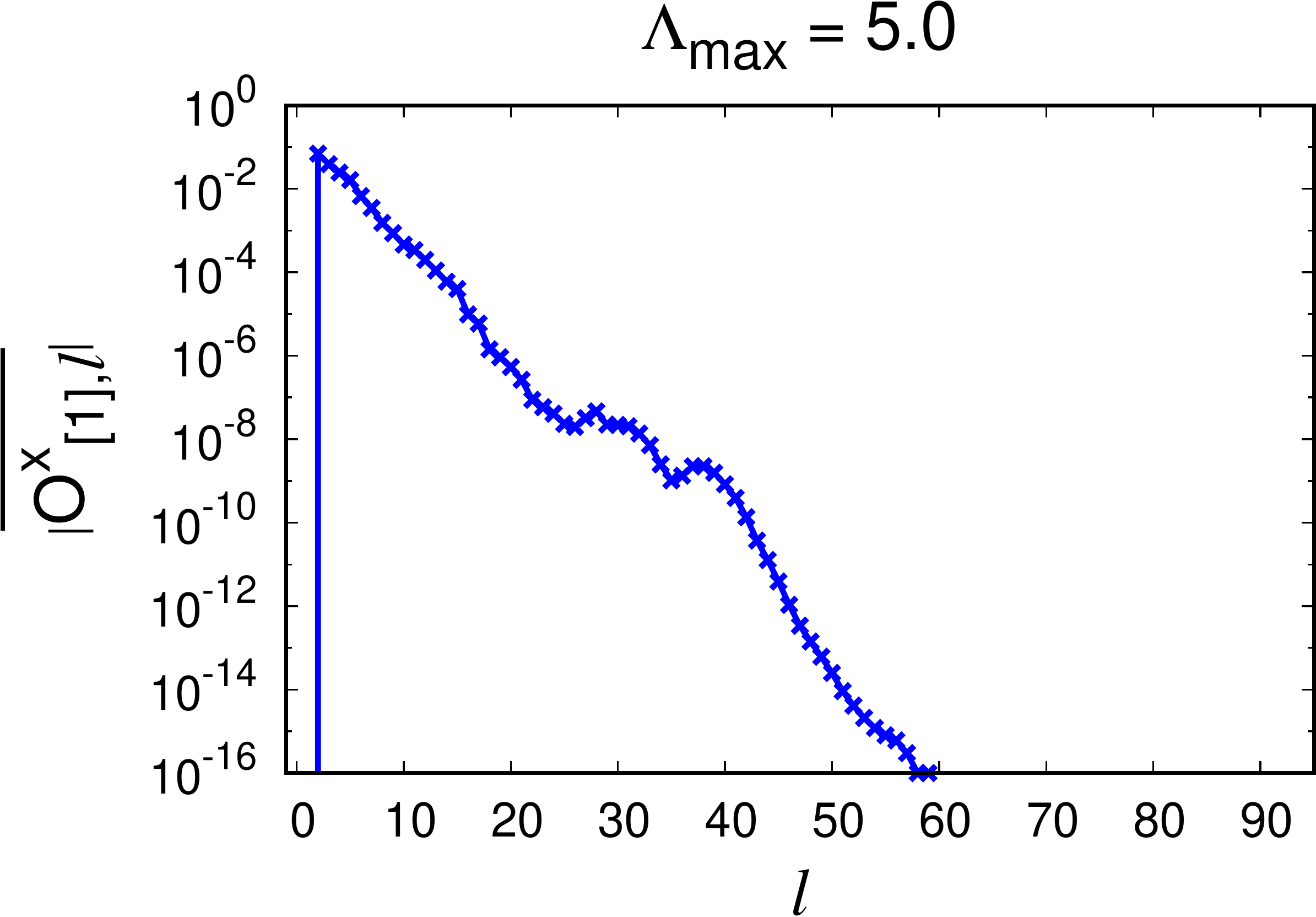}
  \includegraphics[width=0.68\columnwidth]{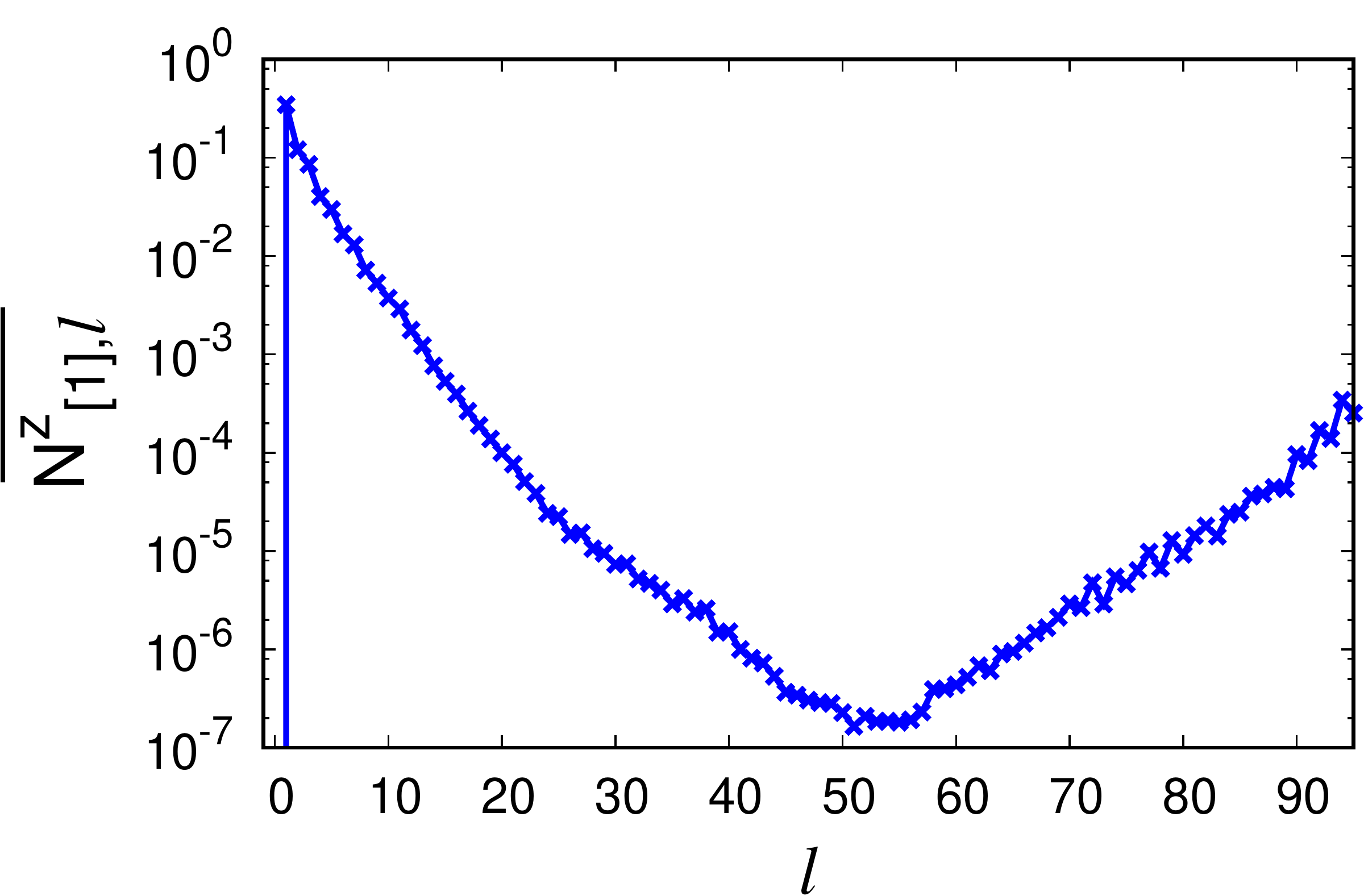}
  \includegraphics[width=0.68\columnwidth]{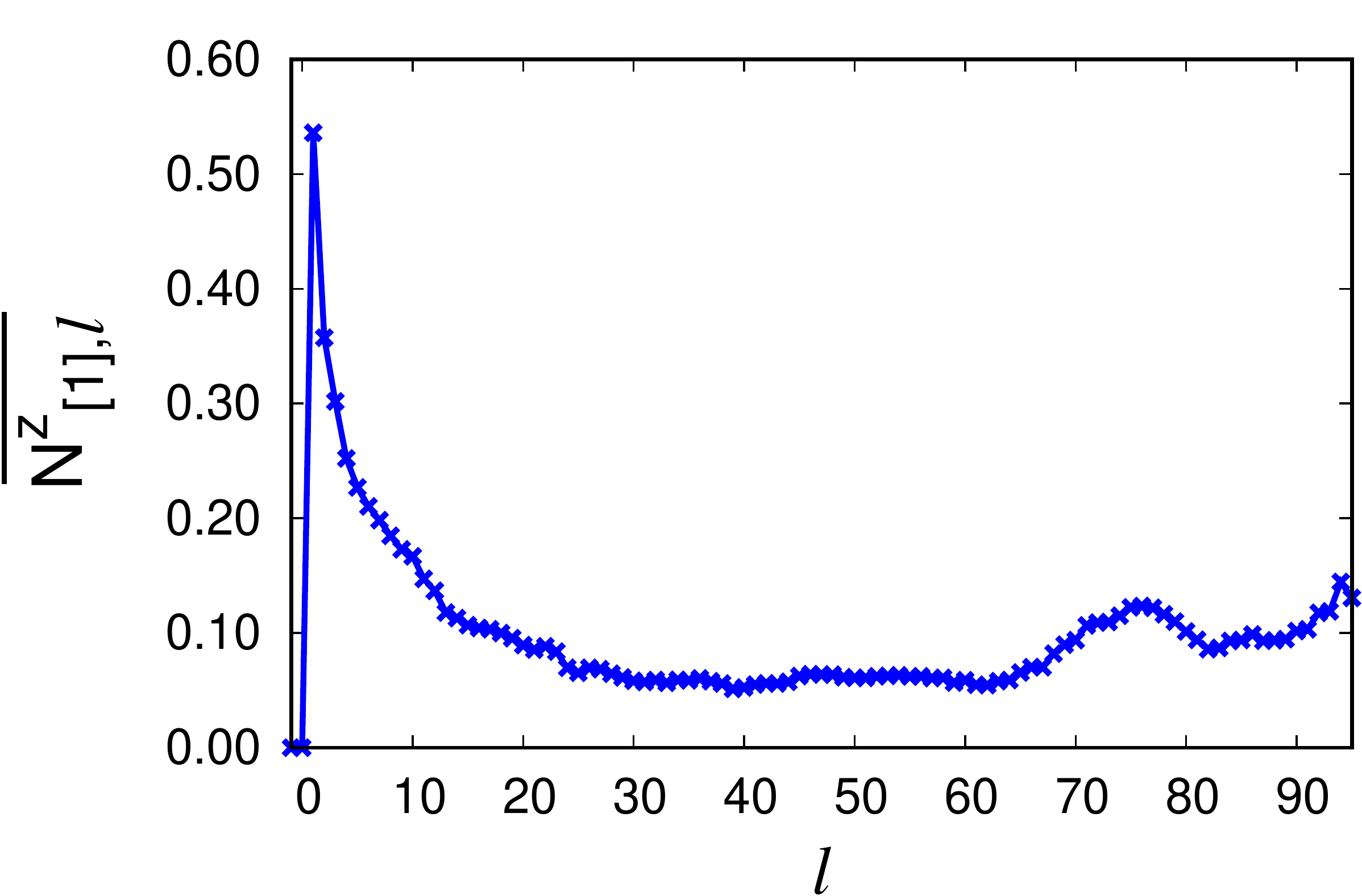}
  \includegraphics[width=0.68\columnwidth]{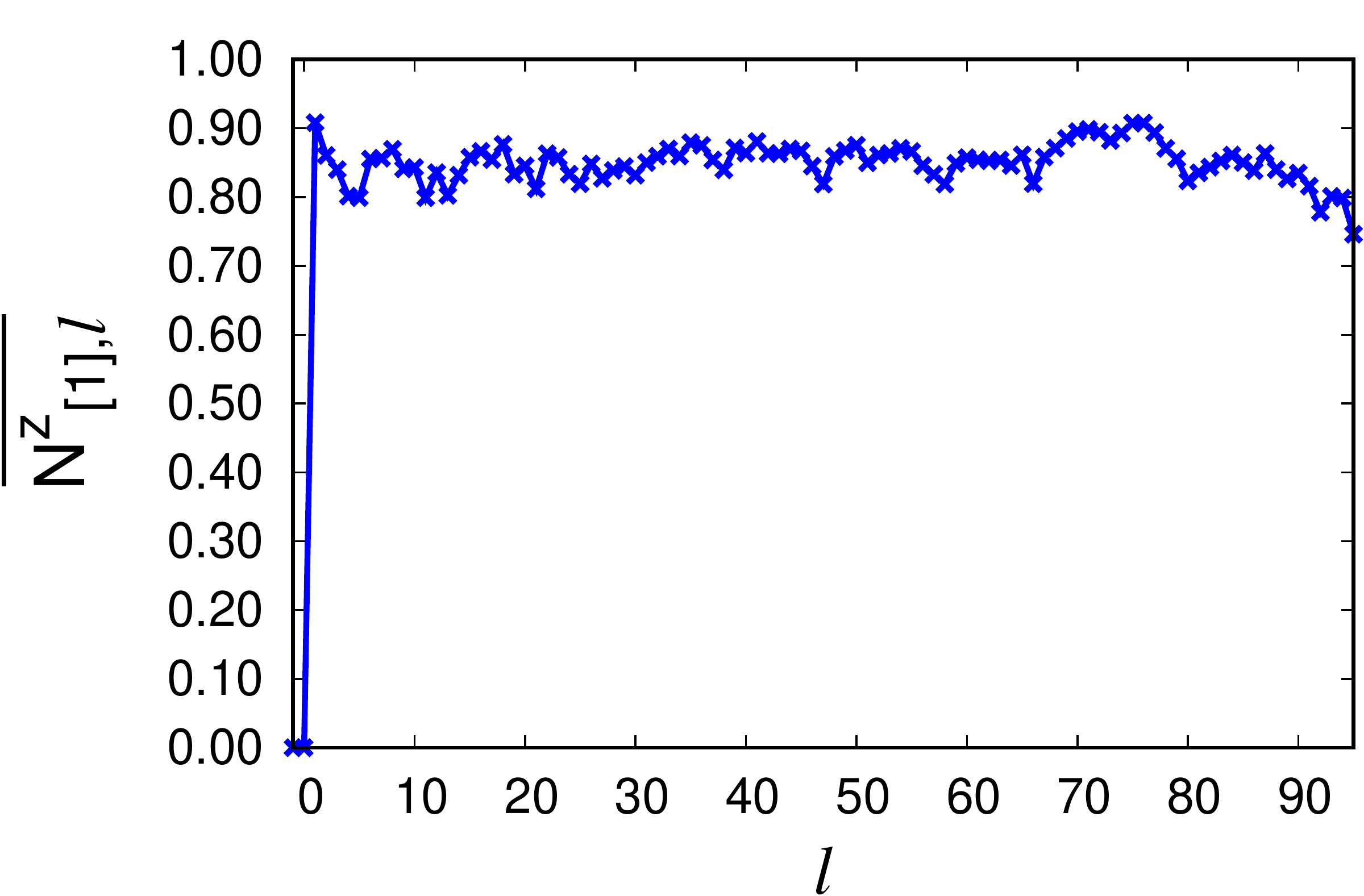}
  \caption{Disorder-averaged string correlator $\overline{|\mathcal{O}^x_{[1],l}|}$ (top panels)
    for ${\Lambda}_{\rm max}=1.5$ (deep Haldane phase, top left panel), ${\Lambda}_{\rm max}=2.3$ (close to the Haldane-N\'eel
    phase transition, top central panel) and ${\Lambda}_{\rm max}=5.0$ (deep N\'eel phase, top right panel),
    and disorder-averaged N\'eel correlator $\overline{\mathcal{N}^z_{[1],l}}$ (bottom panels) for the same values
    of ${\Lambda}_{\rm max}$ as in the top panels. The data refer to the simulation as in Fig.~\ref{fig:phasetransition120sites}.
    The vertical red lines in the top left panel delimit the domain where we compute the spatial averages.
    Apart from spatial fluctuations due to the presence of disorder, the behavior of $\overline{|\mathcal{O}^x_{[1],l}|}$
    and $\overline{\mathcal{N}^z_{[1],l}}$ is in agreement with the expected behavior in the two phases.
    The revival of the $\overline{\mathcal{N}^z_{[1],l}}$ correlator in the bottom left panel, for sufficiently large $l$,
    is due to finite-size effects.}
  \label{fig:disorderaveragedstringneelcorrelators}
\end{figure*}

Because of the presence of the random $\{{\Lambda}_j\}$, the value of $\big\langle \hat{\mathcal{A}}_{\Delta L,\Delta L+l} \big\rangle$, 
as a function of $l$, is expected to fluctuate in space. Thus, the expectation value 
in Eq.~\eqref{eq:spatialaverageofa} is affected by an uncertainty, which we estimate via the standard deviation
\begin{equation}
  \sigma_{\mathcal{A}_h}^2 = \frac{1}{l_2-l_1} \sum_{l=l_1}^{l_2} 
    \left(\langle \hat{\mathcal{A}}_{\Delta L,\Delta L+l}\rangle_h - \mathcal{A}_{{\rm avg},h} \right)^2 \,.
    \label{eq:sigmaspace}
\end{equation}
%
%
%
{The fluctuations over the disorder realizations are in turn computed via the variance computed from the $\{\mathcal{A}_{{\rm avg},h}\}$ in Eq.~\eqref{eq:spatialaverageofa}:}
\begin{equation}
{\sigma^2_{\bar{\mathcal{A}}} = \frac{1}{N_{\rm av}} \sum_{h=1}^{N_{\rm av}} {\left(\mathcal{A}_{{\rm avg},h} -  \overline{\mathcal{A}_{\rm avg}}\, \right)}^2} \, .
\label{eq:sigmadisorder}
\end{equation}
Finally, to motivate the choice of the spatial averages as in Eq.~\eqref{eq:spatialaverageofa},
we show the behavior of the disorder-averaged string correlator, $\overline{|\mathcal{O}^x_{[1],l}|}$,
and the disorder-averaged N\'eel correlator, $\overline{\mathcal{N}^z_{[1],l}}$,
in Fig.~\ref{fig:disorderaveragedstringneelcorrelators}.
Instead of averaging over space for a given realization of disorder, as discussed before, here we average each value
of $|\mathcal{O}^x_{[1],l}|(h)$ and of $\mathcal{N}^z_{[1],l}(h)$, for fixed $k$ and $l$, over $N_{\rm av}$ realization
of disorder, where the symbol ``$(h)$'' indicates that we are computing the expectation values for the $h$-th realization
of disorder [see Eqs.~\eqref{eq:SO-1} and~\eqref{eq:neelcorrelator}]. Explicitly:
\begin{subequations}
\begin{align}
|\overline{\mathcal{O}^x_{[1],l}}| & =\frac{1}{N_{\rm av}}\sum_{h=1}^{N_{\rm av}}|\mathcal{O}^x_{[1],l}|(h)\\
 \overline{\mathcal{N}^z_{[1],l}}  & =\frac{1}{N_{\rm av}}\sum_{h=1}^{N_{\rm av}}\mathcal{N}^z_{[1],l}(h) \,\, .
\end{align}
\label{eq:spacedisorderedoandn}
\end{subequations}
For the plots in Fig.~\ref{fig:disorderaveragedstringneelcorrelators}, we choose $k=0.2\,L$, $L=120$, as in Eq.~\eqref{eq:spatialaverageofa}. The red vertical lines limit the interval of $l$ over which we compute the spatial averages in Eq.~\eqref{eq:spatialaverageofa}. We stress that, as for the computation of the bulk expectation values, the two ways of averaging (average over space/disorder and then average over disorder/space) are actually equivalent, but performing the disorder-average for each value of $k$ and $l$, as in Eqs.~\eqref{eq:spacedisorderedoandn}, allows us to visualize the average spatial behavior of the string and N\'eel correlators.

As we see from Fig.~\ref{fig:disorderaveragedstringneelcorrelators}, apart from spatial fluctuations due to the presence of disorder, the behavior of $\overline{|\mathcal{O}^x_{[1],l}|}$ and $\overline{\mathcal{N}^z_{[1],l}}$
is in agreement with the expected behavior in the two phases (see Sec.~\ref{XXZ:sec}).
We also see that the choice of $l_1=0.3\,L$ and $l_2=0.6\,L$, as in Eq.~\eqref{eq:spatialaverageofa}, allows us to capture the average bulk expectation value, at least sufficiently far away from the transition point. When we are close to the Haldane-N\'eel phase transition (e.g., Fig.~\ref{fig:disorderaveragedstringneelcorrelators}, panels with ${\Lambda}_{\rm max}=2.3$), for finite $L$, our numerical results are affected by finite-size effects, and the correct estimation of the critical point can be then performed only by a finite-size scaling, as explained in Sec.~\ref{XXZ:sec}.

\begin{figure*}[t]
  \includegraphics[width=0.68\columnwidth]{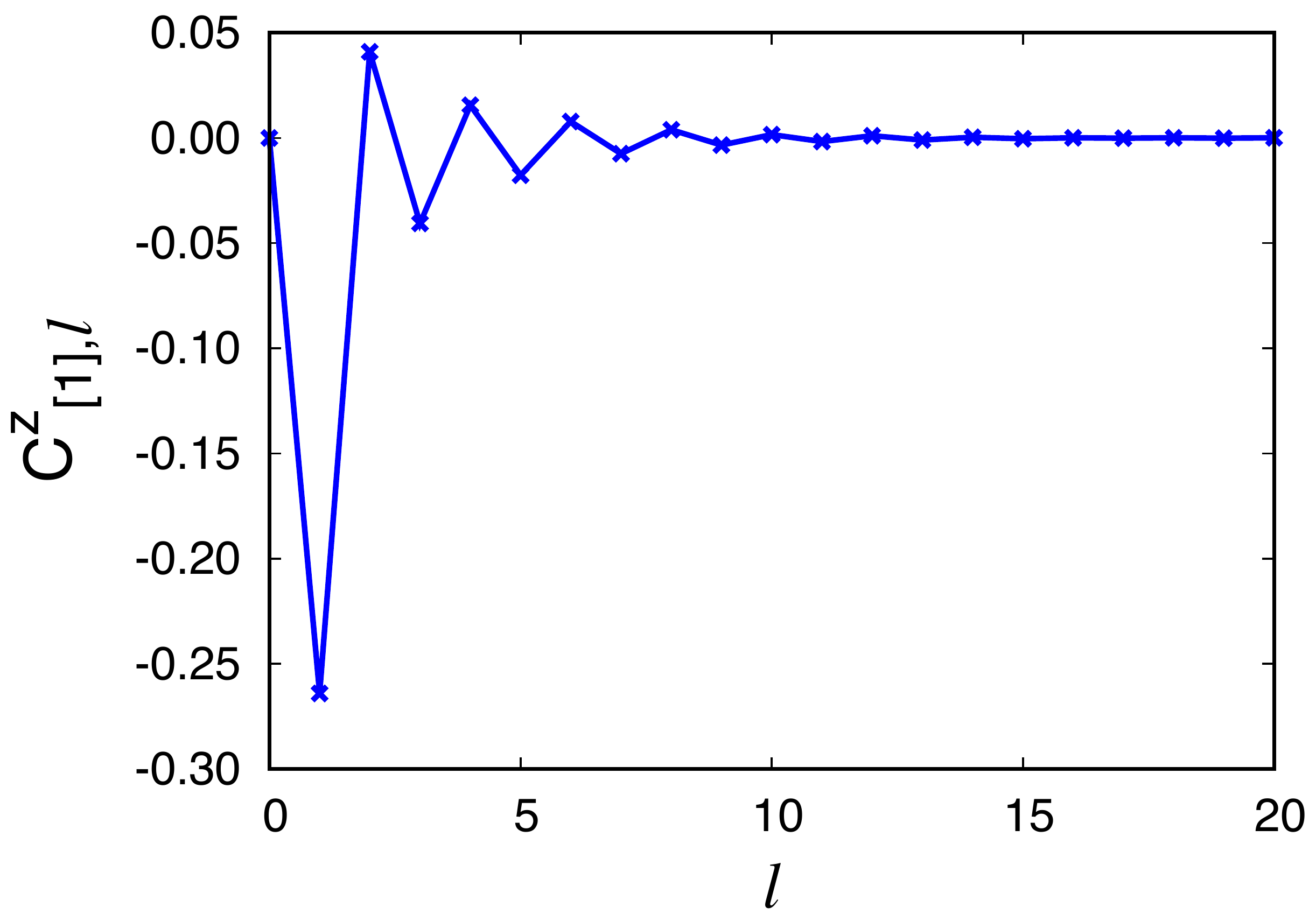}
  \includegraphics[width=0.68\columnwidth]{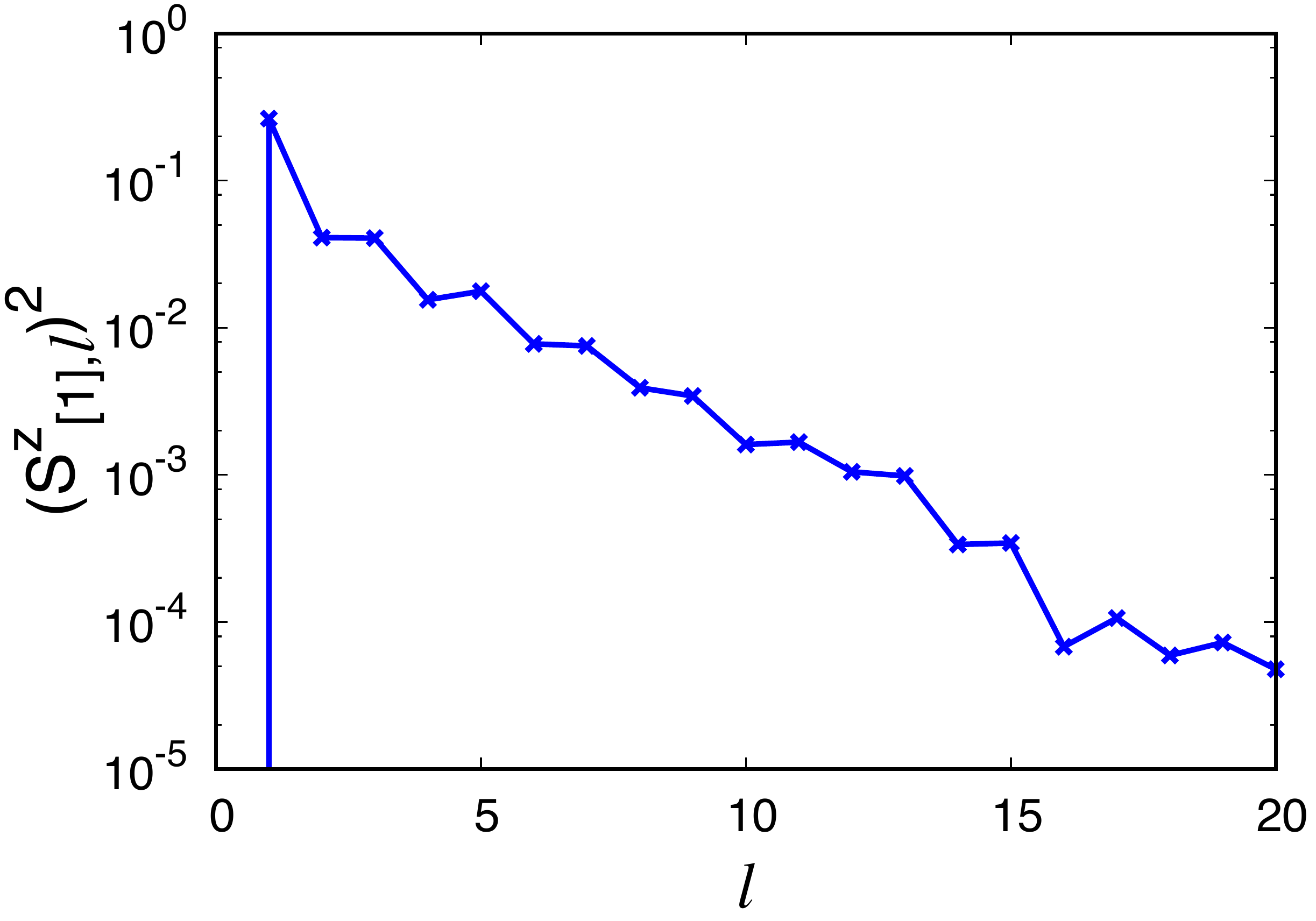}
  \includegraphics[width=0.68\columnwidth]{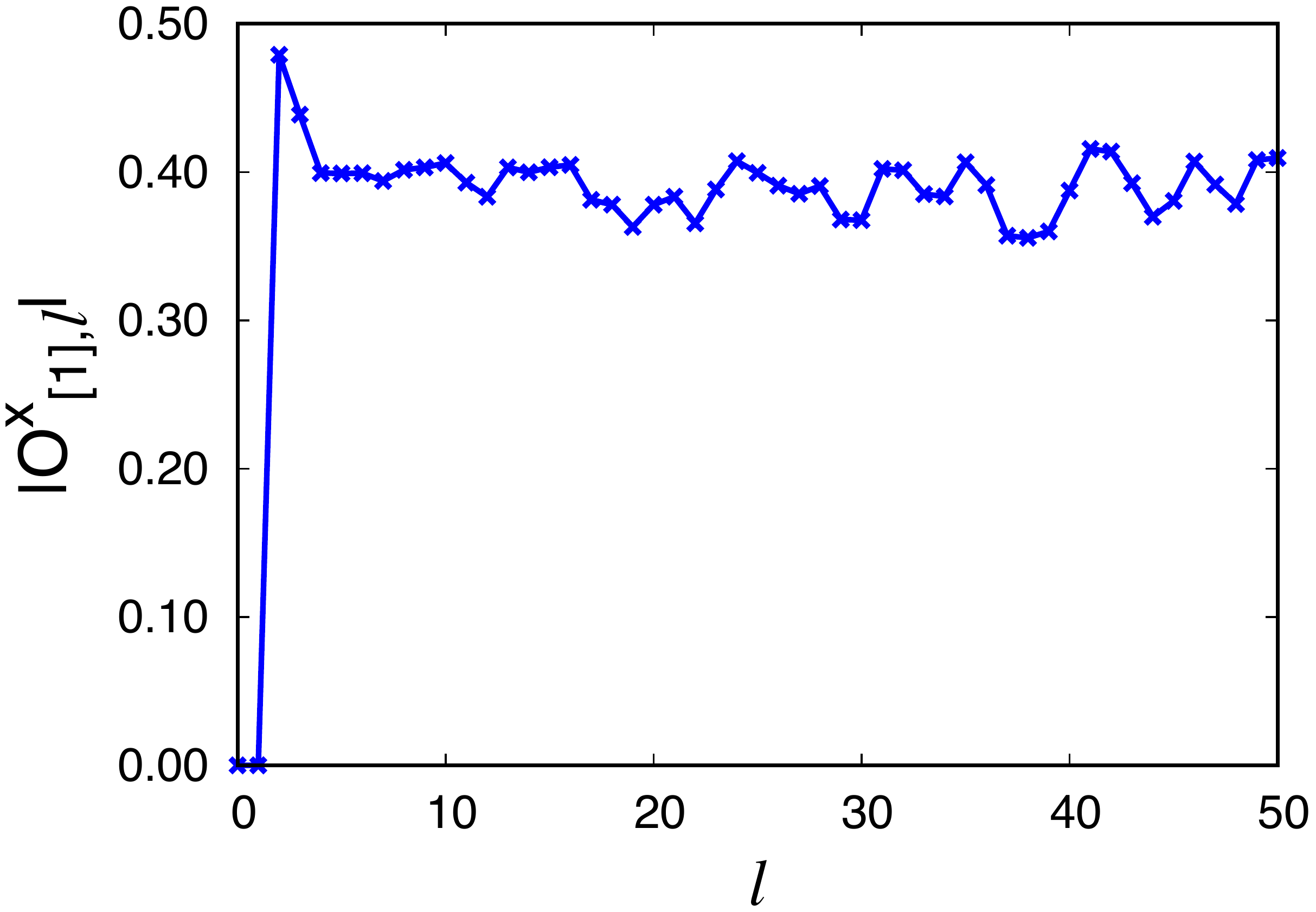}
  \caption{Typical behavior of the two-point correlator $C^z_{[1],l}$ (left panel), 
    the staggered correlator $(\mathcal{S}^z_{[1],l})^2$ (central panel) 
    and the string correlator $|\mathcal{O}^x_{[1],l}|$ (right panel), for a given realization of disorder, 
    in the SO phase at ${\Lambda}_{\rm max}=1.5$. 
    We fix $L=120$ and ${\Lambda}_{\rm min}=0$. 
    Data are shown for different ranges of $l$, in order to highlight the salient properties in the three cases. 
    As explained in the text, we fix $k = \Delta L$, with $\Delta L=0.2\,L$. 
    For ${\Lambda}_{\rm max}=1.5$, the two-point correlator along the $z$-axis oscillates, and it is damped 
    by an exponential decay in the bulk of the chain. This damping is reflected by the exponential decay 
    of the staggered correlator $(\mathcal{S}^z_{[1],l})^2$, whereas the string correlator along the $x$-axis 
    takes a finite value, as expected in the Haldane phase.}
\label{fig:haldanecorrelators}
\end{figure*}

\begin{figure*}[t]
  \includegraphics[width=0.68\columnwidth]{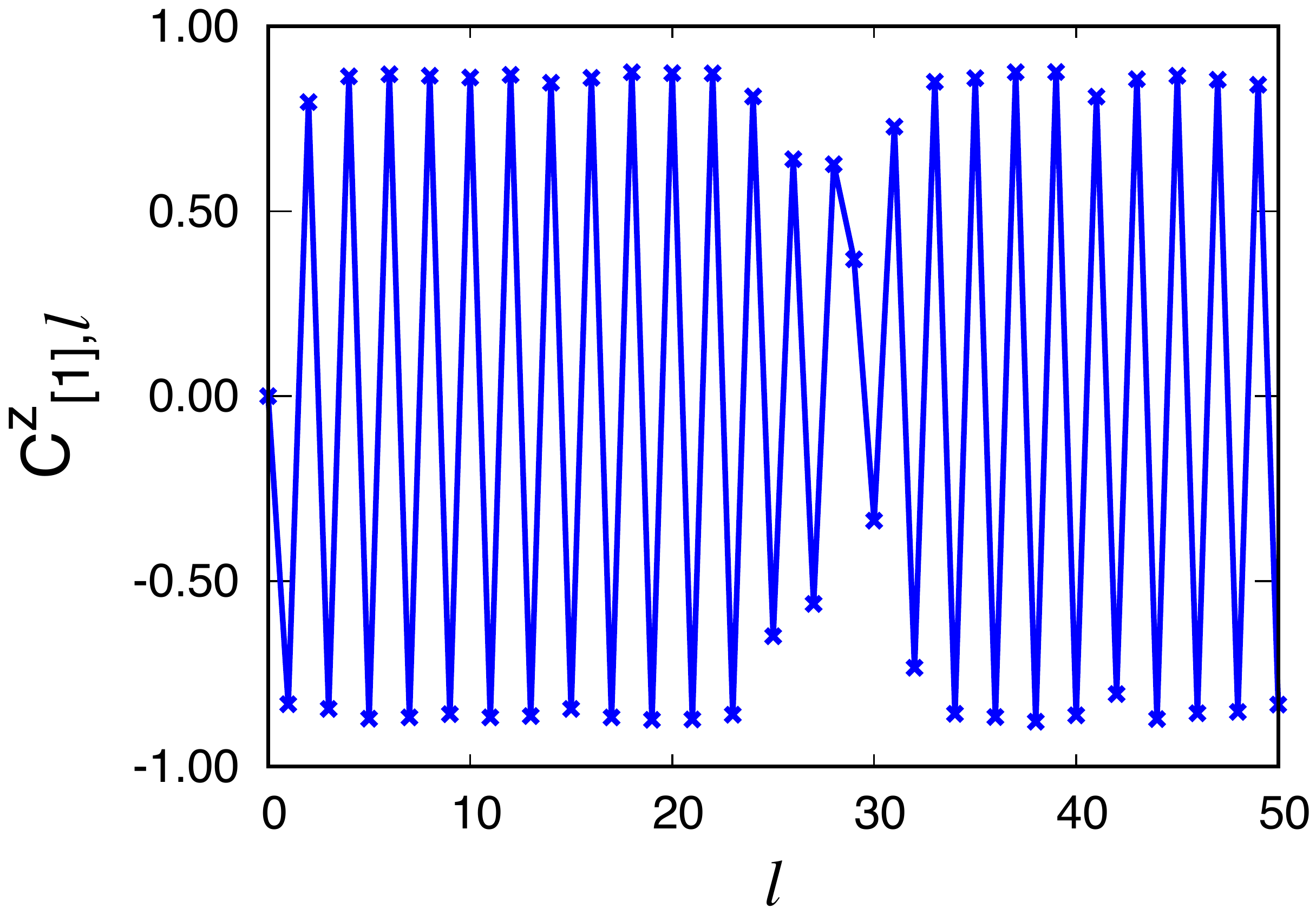}
  \includegraphics[width=0.68\columnwidth]{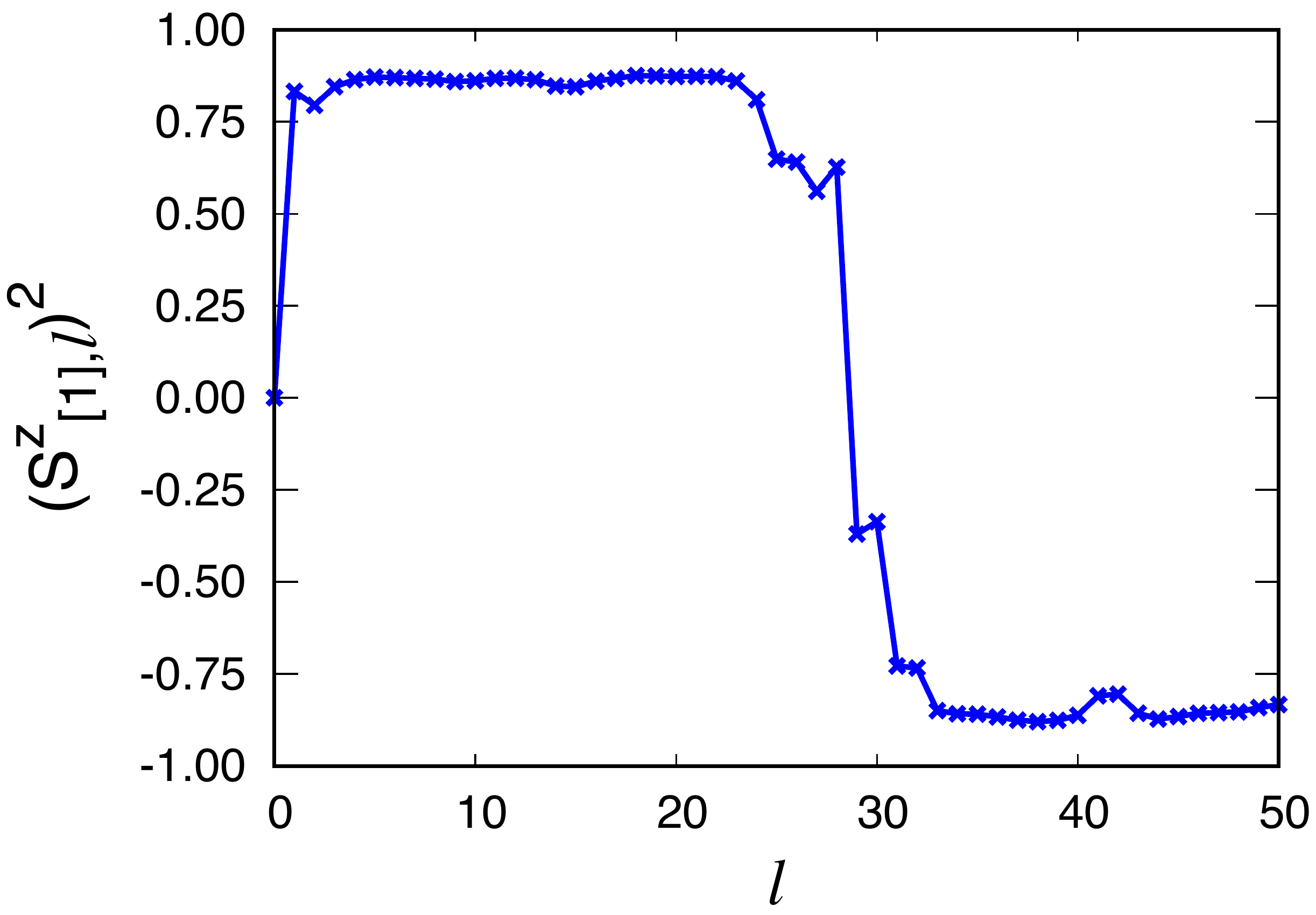}
  \includegraphics[width=0.68\columnwidth]{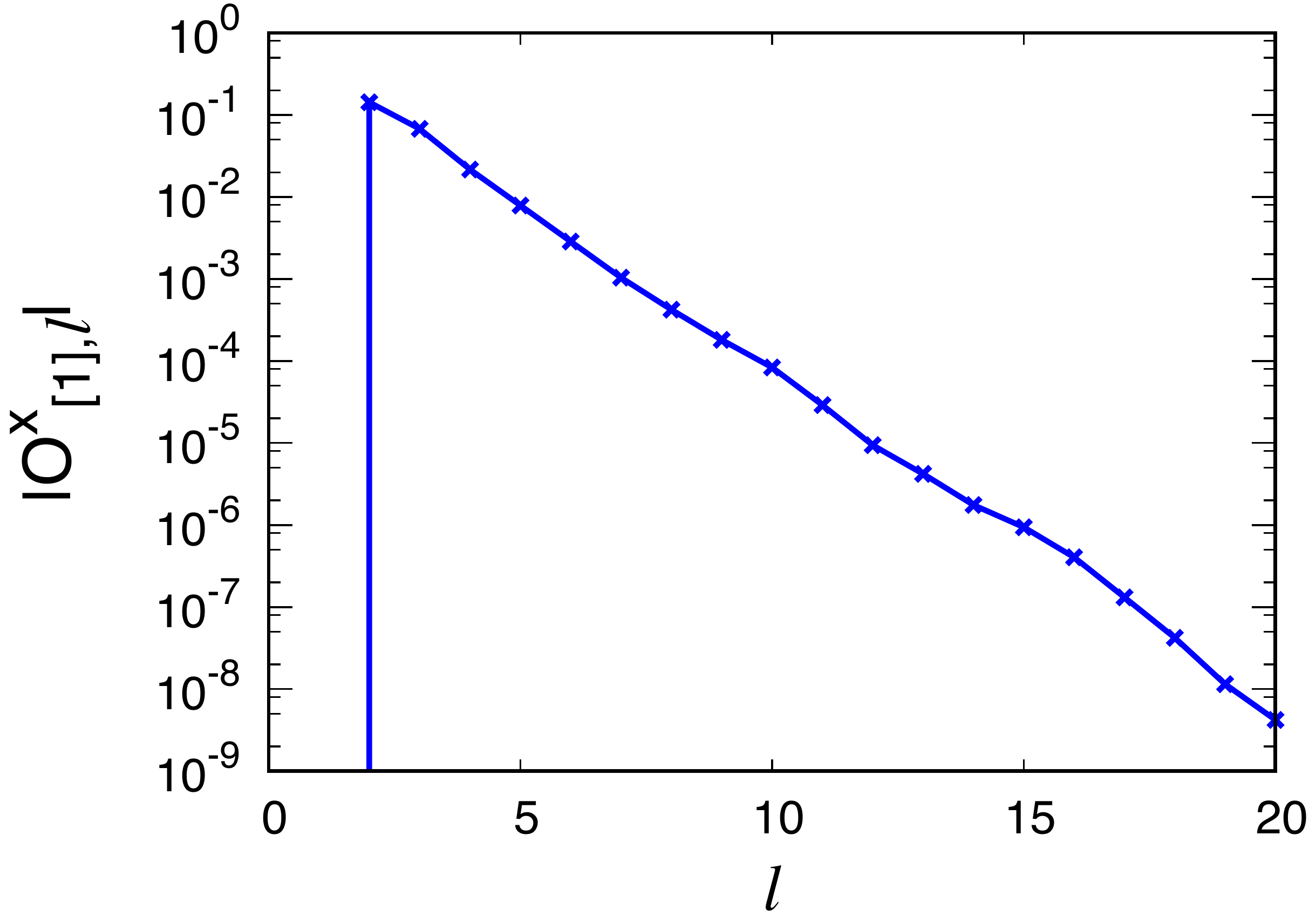}
  \caption{Same analysis as in Fig.~\ref{fig:haldanecorrelators}, but for the AF phase at ${\Lambda}_{\rm max}=5.0$. 
    In this case, the two-point correlator along the $z$-axis oscillates, signaling the presence of AF order. 
    The presence of a kink at $l\simeq 29$ is expected to be a numerical artifact, due to the non-perfect 
    convergence to the GS of the variational algorithm (see main text and Appendix~\ref{sec:appendixdetailsonthekinks}). 
    Such a kink is also seen in the staggered correlator, $(\mathcal{S}^z_{[1],l})^2$, as a sign flip 
    in the staggered pattern. The string correlator along the $x$-axis decays exponentially 
    in the bulk, as expected in the N\'eel phase.}
\label{fig:neelcorrelators}
\end{figure*}

\begin{figure}[!t]
\centering
\includegraphics[width=0.9\columnwidth]{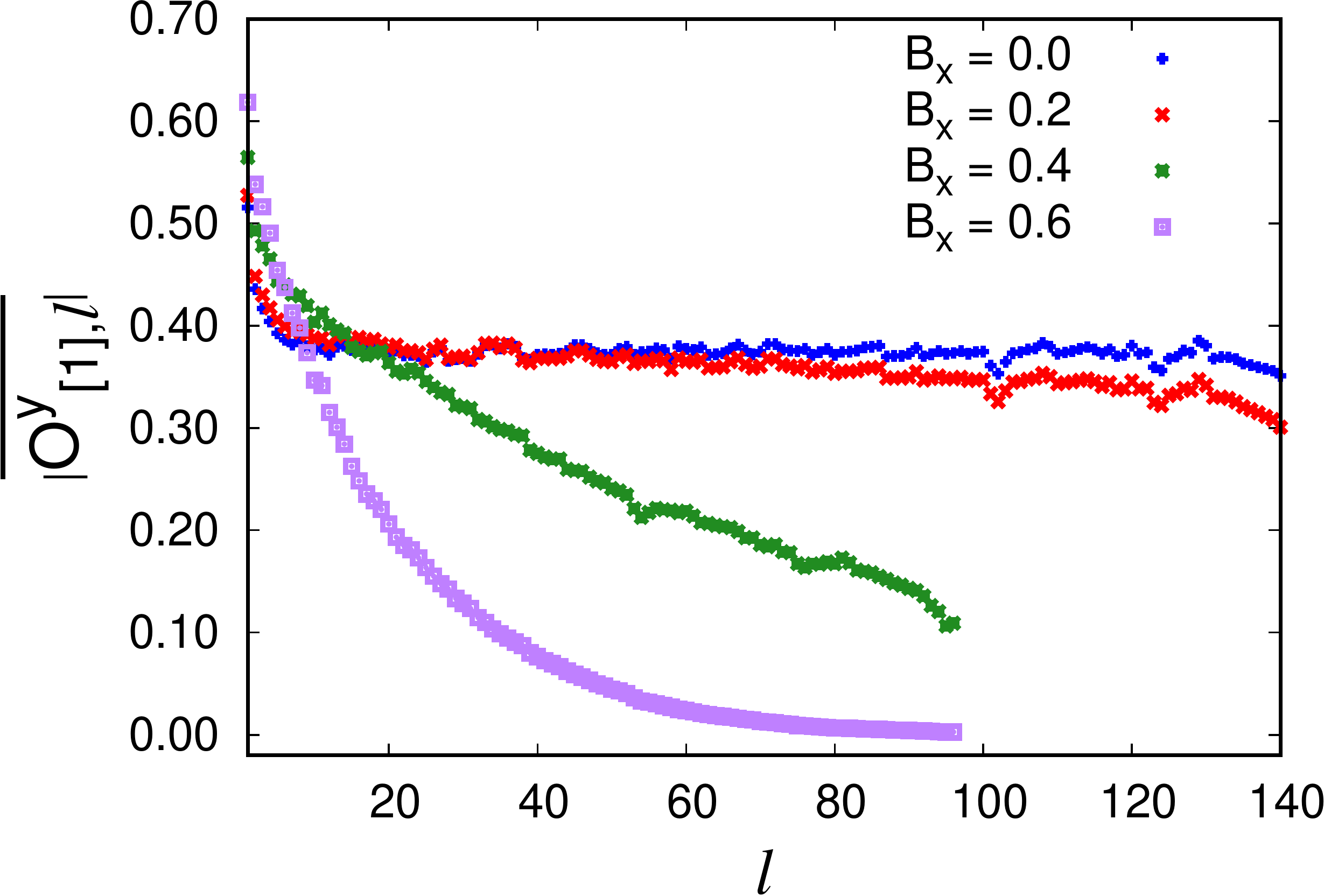}
\caption{Numerical results for the the string correlator $\overline{|\mathcal{O}^y_{[1],l}|}$ as a function of $l$ computed over the GS of the Hamiltonian in Eq.~\eqref{eq:hamiltonianwithmagneticfield}, for different values of $B_x$ as in the legend, using $L=180$ (for $B_x=0,0.2$) and $L=120$ for the others, and $\Lambda_{\rm max}=1.5$. Each point corresponds to the disorder average of $\{|\mathcal{O}^y_{[1],\ell}|(h)\}$, as in Eq.~\eqref{eq:spacedisorderedoandn}. The addition of $B_x\neq0$ makes the string correlator decay to zero in the $l\rightarrow\infty$ limit.}
\label{fig:xxzmodelwithbx}
\end{figure}

\subsection{{Disordered XXZ model with symmetry-breaking magnetic field}}
\label{sec:xxzmodelwithmagneticfield}
{As we mentioned in Section~\ref{Heis:XXZ:subsec}, we can expect the string-ordered phase to be present in the disordered XXZ model as long as the $\mathbb{D}_2$ symmetry is preserved. We now present some numerical data in order to show an example of destruction of the string order in case a symmetry-breaking term is added to the Hamiltonian. Specifically, we simulate the disordered XXZ model in Eq.~\eqref{eq:XXZ} with the inclusion of a magnetic field along the $x$-axis:
\begin{equation}
\hat H_B=\hat H_{\rm XXZ}+B_x\sum_j\hat S^{x}_j \,\, .
\label{eq:hamiltonianwithmagneticfield}
\end{equation}
Since the $\mathbb{D}_2$ symmetry is broken when $B_x\neq0$ in Eq.~\eqref{eq:hamiltonianwithmagneticfield}, we expect SO not to be present in the system~\cite{PerezGarcia_2008,Pollmann_2010}. The result of a simulation with $\Lambda_{\rm min}=0$, $\Lambda_{\rm max}=1.5$ and different values of $B_x$ (in units of $J$) is shown in Fig.~\ref{fig:xxzmodelwithbx}. In order to highlight the different behaviors at long lengths with respect to the $B_x=0$ case, we use $L=180$ for $B_x=0,0.2$ and $L=120$ for $B_x=0.4,0.6$. We compute the disorder average of the string correlator $\overline{|\mathcal{O}^y_{[1],l}|}$ as in Eq.~\eqref{eq:spacedisorderedoandn}. As we see, the addition of $B_x\neq0$ makes the string correlator decay to zero in the $l\rightarrow\infty$ limit, and no SO is present in the system, in agreement with the general arguments presented in Refs.~\onlinecite{PerezGarcia_2008,Pollmann_2010}.}

\subsection{Domain walls}

Let us now consider ${\Lambda}_{\rm min}=0$ and explicitly focus on a case with SO ({\it i}) 
and a case with AF order ({\it ii}).

\indent {\it i)} In Fig.~\ref{fig:haldanecorrelators}, we show the results of a simulation 
with ${\Lambda}_{\rm max}=1.5$ (SO phase) and $L=120$: 
two-point correlator $C^z_{[1],l}\equiv \big\langle \hat S^z_k \hat S^z_{k+l} \big\rangle_h$ (left panel), 
staggered correlator ${(\mathcal{S}^z_{[1],l})}^2$ (central panel), 
and string correlator $|\mathcal{O}^x_{[1],l}|$ (right panel). 
Here all the correlators have been evaluated over a specific realization of disorder, which we term $h$. 
To avoid boundary effects, we choose $\Delta L = 0.2\,L$, $L=120$ (thus, we fix $k=24$). 
For this value of $\Delta_{\rm max}$, we see that $C^z_{[1],l}$ oscillates between positive and negative values, 
for sufficiently small $l$, and it is damped by an exponential decay. 
This is clearly seen in the behavior of $(\mathcal{S}^z_{[1],l})^2$, which exponentially goes to zero 
in the bulk of the chain. On the other hand, the expectation value of the string operator $\mathcal{O}^x_{[1],l}$
takes a finite value in the bulk of the chain.

\indent {\it ii)} We repeat the same analysis as before, but for ${\Lambda}_{\rm max}=5.0$ (AF phase). 
The results are shown in Fig.~\ref{fig:neelcorrelators}. For this value of ${\Lambda}_{\rm max}$, 
the two-point correlator $C^z_{[1],l}$ displays an undamped oscillating pattern, signaling the presence of AF order. 
However, we notice that the pattern reverses at $l\simeq29$, i.e., where the data display a kink (domain wall). 
The presence of such a kink suggests that the AF order appears only locally (the system tends to form domains). 
In order to see if the presence of domain walls in the pattern of the two-point correlator $C^z_{[1],l}$ 
is a physical fact or a numerical artifact, we repeat the simulation $M$ times. We fix the  values of $L$, 
${\Lambda}_{\rm min}$ and ${\Lambda}_{\rm max}$ and the disorder configuration $\{{\Lambda}_j\}$ and in each repetition we vary
the initial random MPS state $|\Psi_{\rm in}\rangle$ at the beginning of the MPS algorithm. 
Our purpose is to verify that different initial random states produce different configurations of kinks 
with different GS energies.

\begin{figure}[t]
  \includegraphics[width=0.85\columnwidth]{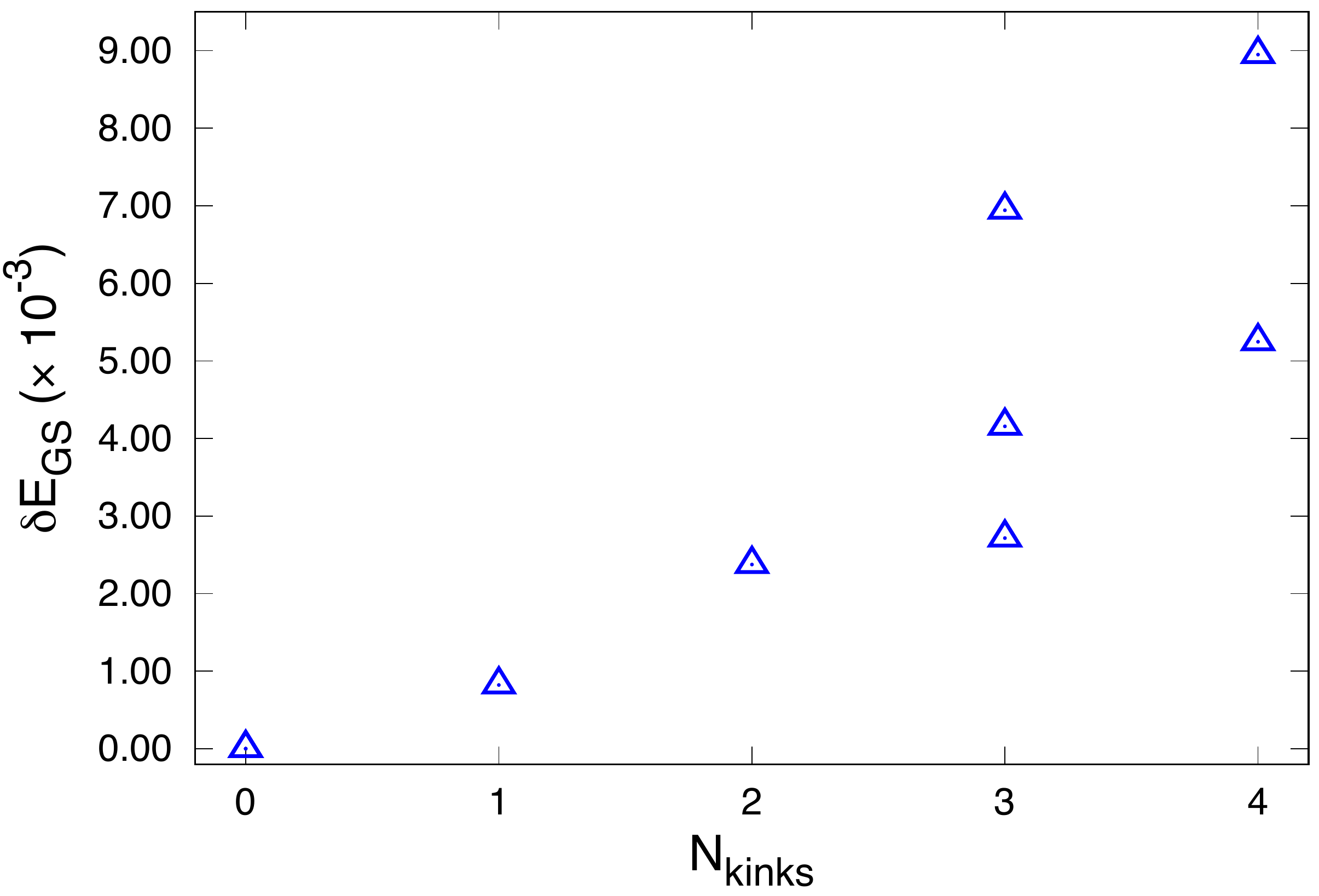}
  \caption{Values of $\delta E_{\rm GS}$ vs.~number of kinks $N_{\rm kinks}$. 
    Each point corresponds to a different choice of the random initial MPS state $|\Psi_{\rm in}(m)\rangle$. 
    Simulations are performed for $L=120$, ${\Lambda}_{\rm min}=0$ and ${\Lambda}_{\rm max}=5.0$, 
    and taking $M=10$ initial states.}
  \label{fig:seedanalysis}
\end{figure}

To give an example, we show in Fig.~\ref{fig:seedanalysis} the result of several simulations 
for $M$ different initial random MPS states: $\big\{ |\Psi_{\rm in}(m)\rangle \big\}_{m = 1 \ldots M}$. 
For each value of $m$, we measure the GS energy,
$E_{\rm GS}(m) = \langle\Psi_{\rm in}(m)|\hat H_{\rm XXZ} |\Psi_{\rm in}(m)\rangle$,
where $\hat H_{\rm XXZ}$ is the Hamiltonian in Eq.~\eqref{eq:XXZ}, and the number of kinks $N_{\rm kinks}(m)$, 
which is obtained from the spatial pattern of $\langle\hat S^z_j\rangle$. 
We define the quantity $\delta E_{\rm GS}(m) = [E_{\rm GS}(m)-{\rm min}_m\{E_{\rm GS}(m)\}]/|{\rm min}_m\{E_{\rm GS}(m)\}|$, 
and compare the values of $\delta E_{\rm GS}(m)$ with the corresponding number of kinks.

As is evident from Fig.~\ref{fig:seedanalysis}, the configurations with zero kinks are associated 
to the lowest value of the GS energy. Furthermore, we see that the GS energy tends to be larger 
for those configurations having a larger number of kinks. In the present case, we have three configurations 
with zero kinks (all associated to the same GS energy), one configuration with one and two kinks, 
three configurations with three kinks, and two configurations with four kinks. 
Simulations ending up with the same number of kinks may have different GS energy, 
since the configuration of kinks along the chain varies as well. 
We see therefore that the number of kinks and their spatial configuration depend on the choice 
of the initial random MPS state, and that the minimum energy is obtained with zero kinks:
We conclude that the presence of domain walls in the magnetization pattern along the chain 
is a numerical artifact due to the fact that the variational MPS algorithm does not perfectly converge 
to the global minimum of the energy functional.

As a consequence, we expect the true GS to have no kinks: also in the presence of disorder, 
there is long-range AF order in the N\'eel phase. This justifies our choice of using 
the N\'eel correlator [Eq.~\eqref{eq:neelcorrelator}] to estimate the staggered correlator 
$\big( \mathcal{S}^z_{[1],l} \big)^2$: the N\'eel correlator does not see the unphysical kinks.


\begin{thebibliography}{100}
\bibitem{Note_winding}
{{In case of periodic boundary conditions the Hamiltonian of the clean system can be written as the direct sum of $2\times2$ blocks, each corresponding to a specific value of the momentum $k$ in the one-dimensional Brillouin zone.
The ground state corresponds to a path in the Bloch sphere parameterized by $k\in[0,2\pi]$. The winding number corresponds to the Berry
phase~\cite{Berry} accumulated in this path, divided by $\pi$. Physically, this number counts how many times the Bloch vector turns around a singularity (a point where the gap of the $2\times 2$ block vanishes), a topological
property which cannot be modified by applying continuous transformations. The only way to modify this property is to make the singularity cross the path,
which implies breaking the continuity of the path (at the crossing point the gap of the $2\times 2$ Hamiltonian closes and the ground-state Bloch vector is not defined). Therefore, when two phases with different
winding number are put into contact, the gap of the Hamiltonian is forced to close at the boundary, that's why the zero energy boundary modes appear.}}
\end{thebibliography}
\end{document}